\documentclass[12pt]{iopart}
\usepackage{iopams}
\usepackage{dsfont}
\usepackage[dvips]{graphicx}

\input cyracc.def 
  \font\tencyr=wncyr10
  \def\cyr{\tencyr\cyracc}

\newcommand{\re}{{\rm e}}
\newcommand{\rd}{{\rm d}}
\newcommand{\half}{\mbox{$\textstyle \frac{1}{2}$}}
\newcommand{\third}{\mbox{$\textstyle \frac{1}{3}$}}
\newcommand{\quat}{\mbox{$\textstyle \frac{1}{4}$}}
\newcommand{\octa}{\mbox{$\textstyle \frac{1}{8}$}}
\newtheorem{prop}{Proposition}

\begin{document}

\title[Information geometry in vapour-liquid
equilibrium]{Information geometry in vapour-liquid equilibrium}

\author{Dorje~C.~Brody${}^1$ and Daniel~W.~Hook${}^2$}

\address{${}^1$Department of Mathematics, Imperial College London,
London SW7 2AZ, UK}

\address{${}^2$Blackett Laboratory,
Imperial College London, London SW7 2AZ, UK}

\begin{abstract}
Using the square-root map $p\to\surd{p}$ a probability density
function $p$ can be represented as a point of the unit sphere
${\mathcal S}$ in the Hilbert space of square-integrable functions.
If the density function depends smoothly on a set of parameters, the
image of the map forms a Riemannian submanifold ${\mathfrak M}
\subset {\mathcal S}$. The metric on ${\mathfrak M}$ induced by the
ambient spherical geometry of ${\mathcal S}$ is the Fisher
information matrix. Statistical properties of the system modelled by
a parametric density function $p$ can then be expressed in terms of
information geometry. An elementary introduction to information
geometry is presented, followed by a precise geometric
characterisation of the family of Gaussian density functions. When
the parametric density function describes the equilibrium state of a
physical system, certain physical characteristics can be identified
with geometric features of the associated information manifold
${\mathfrak M}$. Applying this idea, the properties of vapour-liquid
phase transitions are elucidated in geometrical terms. For an ideal
gas, phase transitions are absent and the geometry of ${\mathfrak
M}$ is flat. In this case, the solutions to the geodesic equations
yield the adiabatic equations of state. For a van der Waals gas, the
associated geometry of ${\mathfrak M}$ is highly nontrivial. The
scalar curvature of ${\mathfrak M}$ diverges along the spinodal
boundary which envelopes the unphysical region in the phase diagram.
The curvature is thus closely related to the stability of the
system.
\end{abstract}

\submitto{\JPA}
%
%

\section{Statistical geometry}
\label{sec:1}

This paper is an overview of the information-geometric description
of vapour-liquid phase transitions in equilibrium statistical
mechanics. The present section begins with a reasonably
self-contained account of the relevant background material on
information geometry. As an illustrative example we shall examine in
some detail the geometry of the space of Gaussian density functions.
The relation between the information measure of Fisher and that of
Shannon and Wiener is also briefly discussed. In later sections
these ideas are applied to the information-geometric
characterisation of the equilibrium properties of noninteracting and
interacting gas molecules. The relevant references are provided in
the bibliographical notes in Section~\ref{sec:4}, where we also
provide a brief and perhaps incomplete history of information
geometry.

\subsection{From probability to geometry}
\label{sec:1.1}

The concept of `information geometry' is a simple one which emerged
from an attempt to discriminate among different probabilities in
statistical analysis. The idea can be sketched as follows. Let
$\{p_i\}_{i=1,2,\ldots,N}$ denote a set of probabilities satisfying
\begin{eqnarray}
0\leq p_i\leq 1 \quad {\rm and}\quad \sum_{i=1}^N p_i = 1.
\label{eq:1.1}
\end{eqnarray}
We introduce the following \textit{square-root map}:
\begin{eqnarray}
p_i\to \xi_i = \sqrt{p_i}. \label{eq:1.2}
\end{eqnarray}
By construction, the square-root probabilities $\{\xi_i\}$ satisfy
the normalisation condition
\begin{eqnarray}
\sum_{i=1}^N \xi_i^2 = 1. \label{eq:1.3}
\end{eqnarray}
If we regard the variables $\{\xi_i\}_{i=1,2,\ldots,N}$ as the
coordinates of a vector in an $N$-dimensional Euclidean space
${\mathds R}^N$, then the normalisation condition (\ref{eq:1.3})
implies that the endpoint of the vector $\{\xi_i\}$ lies on the unit
sphere ${\mathcal S}$ in ${\mathds R}^N$. Now suppose that
$\{\eta_i\}_{i=1,2,\ldots,N}$ corresponds to a second set of
square-root probabilities. Then the vector $\{\eta_i\}$ also lies on
the unit sphere in ${\mathds R}^N$. Hence, we can measure the
\textit{relative separation} or \textit{overlap} of two sets of
probabilities in terms of the angle
\begin{eqnarray}
\phi = \cos^{-1}\sum_{i=1}^N \xi_i \eta_i  \label{eq:1.4}
\end{eqnarray}
between the associated square-root probability vectors. The angular
separation $\phi$ clearly vanishes if $\{\xi_i\}$ and $\{\eta_i\}$
are equal. Conversely, if $\{\xi_i\}$ and $\{\eta_i\}$ are
orthogonal then $\phi$ achieves its maximum value $\frac{1}{2}\pi$.
The angular separation $\phi$ defined in (\ref{eq:1.4}) is known as
the \textit{Bhattacharyya spherical distance}. Note that the cosine
square of the spherical distance resembles the transition
probability in quantum mechanics modelled on a finite-dimensional
Hilbert space.

We turn to the notion of so-called \textit{statistical geometry},
which arises from the embedding of probability density functions in
the Hilbert space of square-integrable functions. In probability
theory one typically deals with a probability density function
$p(x)$ on, say, the real line ${\mathds R}$. For the function
$p:x\to p(x)$ to represent the density of some random variable $X$
we require that $p(x)\geq0$ for all $x\in{\mathds R}$ and that
$\int_{\mathds R}p(x)\rd x=1$. If we consider the \emph{square-root
map}
\begin{eqnarray}
p(x)\to \xi(x) = \sqrt{p(x)}\, , \label{eq:1.5}
\end{eqnarray}
then the function $\xi(x)$ defined in this way belongs to the space
${\cal H}=L^2({\mathds R})$ of square-integrable functions on the
real line. In other words, we \emph{embed} the density functions in
Hilbert space via the square-root map (\ref{eq:1.5}). In particular,
since the square-root density functions satisfy the normalisation
condition
\begin{eqnarray}
\int_{\mathds R}\xi(x)^2\rd x=1, \label{eq:1.5.1}
\end{eqnarray}
the images of the map lie on the unit sphere ${\mathcal
S}\subset{\mathcal H}$.

\begin{figure}
\begin{center}
  \includegraphics[scale=1.0]{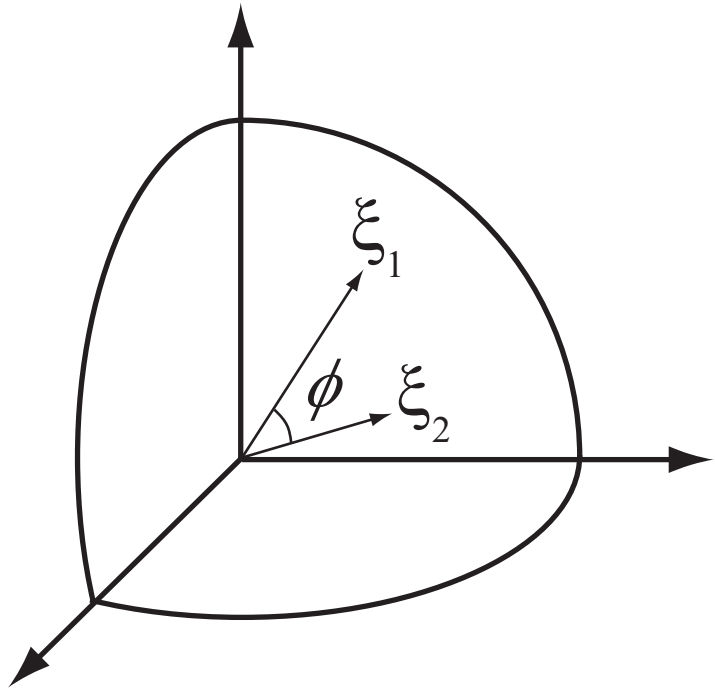}
  \caption{\label{fig:1.0}{\it Bhattacharyya's spherical distance}.
 Two vectors $\xi_1(x)$ and $\xi_2(x)$, corresponding to a pair of
 probability density functions $p_1(x)$ and $p_2(x)$, lie on the
 surface of the positive
 orthant of the unit sphere ${\cal S}$ in Hilbert space. The
 spherical distance between two unit vectors is given by the
 angle $\phi$ defined in equation (\ref{eq:1.7}).
  }
\end{center}
\end{figure}

The advantage of working in Hilbert space ${\mathcal H}$ rather than
the space of density functions is that ${\mathcal H}$ is a vector
space endowed with various geometric features that are familiar from
other branches of physics, such as quantum mechanics or general
relativity.

Suppose we have a pair of density functions $p_1(x)$ and $p_2(x)$,
and  wish to compare the overlap or separation of these two density
functions. If the associated Hilbert space vectors are given
respectively by $\xi_1(x)$ and $\xi_2(x)$, then the overlap is
measured in terms of the inner product $\int_{\mathds R} \xi_1(x)
\xi_2(x)\rd x$. Since $\|\xi_1\|=\|\xi_2\|=1$, i.e. both vectors
have unit norm, this overlap is given by the cosine of the angular
separation. It follows that the Bhattacharyya spherical distance
between two square-root density functions is
\begin{eqnarray}
\phi = \cos^{-1}\int_{\mathds R} \xi_1(x) \xi_2(x)\rd x .
\label{eq:1.7}
\end{eqnarray}
This idea is illustrated schematically in Fig.~\ref{fig:1.0}.

\subsection{Parametric density and Fisher-Rao geometry}
\label{sec:1.2}

In theoretical statistics one typically deals with a parametric
family of probability density functions $p_\theta(x) =p(x|\theta)$.
Here $\theta$ denotes one or more real parameters. For example, a
Gaussian density function is characterised by two parameters, i.e.
the mean $\mu$ and the variance $\sigma^2$. For each value, or set
of values, of $\theta$ we have the normalisation condition
$\int_{\mathds R} p_\theta(x)\rd x=1$.

In problems of statistical inference, it is often convenient to
consider the \textit{log-likelihood function}
\begin{eqnarray}
l_{\theta}(x)=\ln p(x|\theta). \label{eq:1.1.8}
\end{eqnarray}
However, in physics it is more natural to work with the square-root
density function
\begin{eqnarray}
\xi_{\theta}(x) = \sqrt{p(x|\theta)}, \label{eq:1.1.9}
\end{eqnarray}
since, as indicated above, this permits formulation of the problem
in a real Hilbert-space context. As before, for each given $\theta$
the density function is mapped to a point on the unit sphere
${\mathcal S}\subset{\cal H}$ by the prescription (\ref{eq:1.1.9}).
If the value of $\theta$ is changed, the image under the map in
general also varies on ${\mathcal S}$. We assume that the density
function is at least twice differentiable with respect to the
parameters. Then as the parameters change continuously, the image
point on ${\mathcal S}$ will vary smoothly over a parametric
subspace ${\mathfrak M}$ of the sphere ${\mathcal S}$.

Given a parametric subspace ${\mathfrak M}\subset{\mathcal S}$ the
metric of the ambient sphere ${\mathcal S}$ induces a Riemannian
metric on the subspace in the usual way. This can be seen as
follows. Recall that the Hilbert space inner product is defined by
\begin{eqnarray}
\langle \xi,\eta\rangle = \int_{\mathds R} \xi(x) \eta(x) \rd x.
\label{eq:1.11}
\end{eqnarray}
Therefore, if we set
\begin{eqnarray}
\xi(x)=\xi_\theta(x)\quad {\rm and}\quad \eta(x)= \xi_\theta(x) +
\partial_i \xi_\theta(x)\rd\theta^i,
\end{eqnarray}
where $\partial_i = \partial/\partial \theta^i$, then the squared
distance $\rd s^2$ of the difference vector $\xi(x)-\eta(x)$ is
given by
\begin{eqnarray}
\rd s^2 = \left(\int_{\mathds R} \partial_i \xi_\theta(x) \partial_j
\xi_\theta(x) \rd x \right) \rd \theta^i \rd \theta^j.
\label{eq:1.12}
\end{eqnarray}

Before we proceed further with the derivation of the metric, let us
introduce the statistical notion of the \textit{Fisher information
matrix}, which is usually defined by
\begin{eqnarray}
{G}_{ij} = \int_{\mathds R} p_\theta(x)\, \partial_i l_\theta(x) \,
\partial_j l_\theta(x) \, \rd x , \label{eq:1.13}
\end{eqnarray}
where $l_{\theta}(x)$ is the log-likelihood density
(\ref{eq:1.1.8}). The Fisher information matrix is important in
statistics because it provides a lower bound for the variance of a
parameter estimate.Consider, for example, the case of a
one-parameter family of density functions. That is, we have a
density function $p_\theta(x)$ that depends upon a single unknown
parameter $\theta$. The objective is thus to estimate the parameter
by performing observations. If $T(x)$ is an unbiased estimator for
$\theta$, i.e. if the expectation of $T(x)$ with respect to
$p_\theta(x)$ yields $\theta$, then we have
\begin{eqnarray}
\int_{\mathds R} (T(x)-\theta) \xi_\theta(x)^2\, \rd x = 0.
\end{eqnarray}
Differentiating with respect to $\theta$ we obtain
\begin{eqnarray}
\int_{\mathds R} (T(x)-\theta)\, \xi_\theta(x)\, \partial_\theta
\xi_\theta(x)\, \rd x = \half,
\end{eqnarray}
where $\partial_\theta=\partial/\partial\theta$. By the Schwarz
inequality
\begin{eqnarray}
\fl \left(\int_{\mathds R} (T(x)-\theta)\, \xi_\theta(x)\,
\partial_\theta \xi_\theta(x)\, \rd x \right)^2 \leq \left(
\int_{\mathds R} (T(x)-\theta)^2 \xi_\theta(x)^2\, \rd x\right)
\left( \int_{\mathds R} (\partial_\theta\xi_\theta(x))^2\, \rd x
\right) \label{eq:1.16a}
\end{eqnarray}
we thus find
\begin{eqnarray}
\int_{\mathds R} (T(x)-\theta)^2 \xi_\theta(x)^2\, \rd x \geq
\frac{1}{4\int_{\mathds R} (\partial_\theta\xi_\theta(x))^2\, \rd
x}. \label{eq:1.15}
\end{eqnarray}
Note that the left side is the variance of the estimator, whereas
the denominator of the right side is the one-parameter form of the
Fisher information matrix. The relation (\ref{eq:1.15}) provides a
lower bound for the variance, and is known as the
\textit{Cram\'er-Rao inequality}. The inequality in (\ref{eq:1.16a})
is attained only when the two vectors are proportional, that is,
$\partial_\theta \xi_\theta(x) = c(T(x)-\theta)\, \xi_\theta(x)$ for
some constant $c$. By scaling $\theta$ we can set $c=\frac{1}{2}$
without loss of generality. Hence, the lower bound of the variance
is attained only if the square-root density assumes an exponential
form:
\begin{eqnarray}
\xi_\theta(x) = \frac{\exp\left( \half \theta T(x)\right)}
{\left(\int_{\mathds R} \exp\left(\theta T(x)\right)\rd x
\right)^{1/2}} . \label{eq:1.18}
\end{eqnarray}
The exponential family (\ref{eq:1.18}) plays an important role in
the applications to statistical mechanics considered below.

In a multi-parameter context the reciprocal of the Fisher
information matrix determines lower bounds for the variance in an
analogous manner. From the geometrical viewpoint, the significance
of the Fisher information matrix is that it defines the induced
Riemannian metric on the parametric subspace ${\mathfrak M}$ of the
unit sphere ${\mathcal S}$ in ${\mathcal H}$. Specifically,
comparing (\ref{eq:1.12}) and (\ref{eq:1.13}) we see that
\begin{eqnarray}
\rd s^2 = \quat\, {G}_{ij} \rd\theta^i \rd\theta^j. \label{eq:1.19}
\end{eqnarray}
The metric $\frac{1}{4}\,{G}_{ij}$ on ${\mathfrak M}$ will be
referred to as the \textit{Fisher-Rao metric}. The factor of a
quarter is purely a matter of convention, and the Fisher-Rao metric
is thus given by a quarter of the Fisher information matrix.

\subsection{Riemannian structure of the exponential family}
\label{sec:1.3}

We introduce here some elementary concepts in Riemannian geometry
that are relevant to the ensuring discussion. We note first that all
equilibrium distributions that we consider here are represented in
the exponential form
\begin{eqnarray}
p_\theta(x) = q(x) \exp\left( -\sum_i \theta^iH_i(x)-\psi(\theta)
\right), \label{eq:3.1}
\end{eqnarray}
where $\{\theta^i\}_{i=1,2,\ldots}$ are parameters, $q(x)$
represents the prescribed equilibrium state at $\theta^i=0$ for all
$i$, and the functions $\{H_i(x)\}_{i=1,2,\ldots}$ determine the
form of the energy. In other words, we shall only consider
equilibrium states represented in the \textit{canonical} form. The
parameters $\{\theta^i\}$ may include inverse temperature, chemical
potential, pressure, magnetic field, and so on, whereas the
functions $H_i(x)$ may represent system energy, particle number,
system volume, magnetisation, and so on. The variable $x$ ranges
over the phase space ${\boldsymbol{\Gamma}}$ of the system. The
function
\begin{eqnarray}
\psi(\theta) = \ln \int_{\boldsymbol{\Gamma}} \exp\left( -\sum_i
\theta^iH_i(x) \right)q(x) \rd x \label{eq:3.2}
\end{eqnarray}
determines the overall normalisation. We refer to $\psi(\theta)$ as
the \textit{thermodynamic potential} of the system. It should be
evident by inspection that
\begin{eqnarray}\nonumber
-\frac{\partial \psi}{\partial
\theta^i}&=&\frac{\int_{\boldsymbol{\Gamma}} H_i(x) \exp\left(
-\sum_i \theta^i H_i(x)\right) q(x) \rd x}
{\int_{\boldsymbol{\Gamma}}\exp\left(-\sum_i \theta^i H_i(x)
\right)q(x) \rd x} \\ &=&\int_{\boldsymbol{\Gamma}} H_i(x)
p_\theta(x)\rd x.\label{eq:3.2b}
\end{eqnarray}
In other words, the first derivative of $\psi(\theta)$ with respect
to $\theta^i$ determines the expectation value of $H_i(x)$ in the
equilibrium state (\ref{eq:3.1}). As we shall indicate below,
analogous calculations show that higher derivatives of the
thermodynamic potential $\psi(\theta)$ determine higher moments of
the functions $\{H_i(x)\}$.

If the equilibrium density function assumes the form (\ref{eq:3.1}),
then the expressions for the corresponding Fisher-Rao metric and the
coefficients of the associated metric connection on the statistical
manifold ${\mathfrak M}$ are simplified. Let us state the results
first.

\begin{prop}
For a density function of the exponential form {\rm(\ref{eq:3.1})}
the Fisher-Rao metric ${G}_{ij}$ and the Christoffel symbols
$\Gamma_{ijk}={G}_{il}\Gamma^l_{\,jk}$ are given, respectively, by
\begin{eqnarray}
{G}_{ij} = \partial_i\partial_j \psi(\theta) \label{eq:1.23}
\end{eqnarray}
and
\begin{eqnarray}
\Gamma_{ijk}=\half\,\partial_i\partial_j\partial_k \psi(\theta),
\label{eq:1.24}
\end{eqnarray}
in terms of the canonical parametrisation $\{\theta^i\}$ on
${\mathfrak M}$, where $\partial_i=\partial/\partial \theta_i$.
\label{prop:1}
\end{prop}

The Christoffel symbols characterise the geodesics on ${\mathfrak
M}$. Specifically, to find the shortest path from $a$ to $b$ on
${\mathfrak M}$ we consider the variational problem:
\begin{eqnarray}
\delta \int_a^b \rd s = 0. \label{eq:1.25}
\end{eqnarray}
From (\ref{eq:1.19}) we find
\begin{eqnarray}
4\delta \rd s^2 = \rd\theta^i \rd\theta^k \frac{\partial
G_{ik}}{\partial \theta^l}\delta\theta^l +2{G}_{ik}\rd \theta^i
\rd\delta\theta^k . \label{eq:1.26}
\end{eqnarray}
Bearing in mind that the right side of (\ref{eq:1.26}) equals $8\rd
s\,\delta \rd s$, we obtain
\begin{eqnarray}
\int_a^b \left[ \half \frac{\rd\theta^i}{\rd s}
\frac{\rd\theta^k}{\rd s} \frac{\partial{G}_{ik}}{\partial
\theta^k}\, \delta\theta^l + {G}_{ik} \frac{\rd\theta^i}{\rd s}
\frac{\rd\delta\theta^k}{\rd s} \right]\rd s = 0. \label{eq:1.27}
\end{eqnarray}
Integrating the second term in the integrand by parts and writing
$u^i=\rd\theta^i/\rd s$ we see that (\ref{eq:1.27}) reduces to
\begin{eqnarray}
\int_a^b \left[ \half u^i u^k \frac{\partial{G}_{ik}} {\partial
\theta^l} - \frac{\rd}{\rd s} \left({G}_{il} u^i \right) \right]
\delta\theta^l \rd s = 0. \label{eq:1.28}
\end{eqnarray}
Since this must hold for arbitrary $\delta\theta^l$ we have
\begin{eqnarray}
\half u^i u^k \frac{\partial{G}_{ik}}{\partial \theta^l} -
\frac{\rd}{\rd s} \left({G}_{il} u^i \right) = 0. \label{eq:1.29}
\end{eqnarray}
Writing
\begin{eqnarray}
\Gamma^i_{\,kl} = \half {G}^{im} \left( \frac{\partial
G_{mk}}{\partial\theta^l}+\frac{\partial{G}_{ml}} {\partial
\theta^k}-\frac{\partial{G}_{kl}}{\partial\theta^m} \right)
\label{eq:1.30}
\end{eqnarray}
for the Christoffel symbol we find that (\ref{eq:1.29}) can be
expressed in the form
\begin{eqnarray}
\frac{\rd^2\theta^i}{\rd s^2}+ \Gamma^i_{\,kl}
\frac{\rd\theta^k}{\rd s}\frac{\rd\theta^l}{\rd s} = 0.
\label{eq:1.31}
\end{eqnarray}
This is the \textit{geodesic equation} that determines the shortest
paths on ${\mathfrak M}$. Owing to their nonlinearity, geodesic
equations do not generally admit elementary analytic solutions,
although in some cases one can solve (\ref{eq:1.31}) in closed form,
as in the Gaussian example discussed below.

\vspace{0.2cm} \textit{Proof of Proposition~\ref{prop:1}}. From
(\ref{eq:3.1}) we have
\begin{eqnarray}
\partial_i l_\theta(x)=-\left( H_i(x)+\partial_i\psi(\theta)
\right). \label{eq:1.32}
\end{eqnarray}
On the other hand, differentiating the normalisation condition
\begin{eqnarray}
\int_{\boldsymbol{\Gamma}} p_\theta(x)\rd x=1
\end{eqnarray}
once with respect to $\theta^i$ and using $\partial_i p_\theta(x)=
p_\theta(x) \partial_i l_\theta(x)$ we obtain
\begin{eqnarray}
\int_{\boldsymbol{\Gamma}} p_\theta(x) \left(
H_i(x)+\partial_i\psi(\theta) \right)\rd x = 0, \label{eq:3.6}
\end{eqnarray}
whence it follows that the expectation of $H_i(x)$ with respect to
$p_\theta(x)$ is given by $-\partial_i\psi(\theta)$, as shown in
(\ref{eq:3.2b}). Differentiating (\ref{eq:3.6}) with respect to
$\theta^j$, we find that
\begin{eqnarray}
-\int_{\boldsymbol{\Gamma}} p_\theta(x) \left(
H_i(x)+\partial_i\psi(\theta) \right) \left(
H_j(x)+\partial_j\psi(\theta) \right)\rd x + \partial_i\partial_j
\psi(\theta) = 0. \label{eq:3.7}
\end{eqnarray}
In view of (\ref{eq:1.32}) and (\ref{eq:1.13}), this implies that
the Fisher-Rao metric ${G}_{ij}$ is given by (\ref{eq:1.23}). From
(\ref{eq:1.30}) we have
\begin{eqnarray}
\Gamma_{ikl} = \half \left( \partial_l{G}_{ik}+ \partial_k
G_{il}-\partial_i{G}_{kl} \right), \label{eq:1.36}
\end{eqnarray}
but since the metric is given by (\ref{eq:1.23}) we immediately
deduce the expression (\ref{eq:1.24}) for the Christoffel symbols.
\hspace*{\fill} $\square$ \vspace{0.2cm}

It is important to note that if we choose an alternative
parametrisation for ${\mathfrak M}$, then the components of the
metric tensor and the Christoffel symbol cannot be calculated using
the simple expressions given in Proposition~\ref{prop:1}, and we
must use the defining equations (\ref{eq:1.13}) and (\ref{eq:1.30})
to determine these quantities. Also note that the metric tensor is
the covariance matrix of the functions $\{H_i(x)\}$, whereas the
components of $\Gamma_{ijk}$ are third-order cross-moments of
$\{H_i(x)\}$.

In terms of the Christoffel symbols $\Gamma^i_{\,jk}$ the Riemann
curvature tensor $R^i_{\ ljk}$ can be expressed as
\begin{eqnarray}
R^i_{\ ljk} = \partial_k\Gamma^i_{\,lj}-\partial_j\Gamma^i_{\,lk}-
\Gamma^i_{\,jh}\Gamma^h_{\,lk}+\Gamma^i_{\,kh}\Gamma^h_{\,lj}.
\label{eq:3.8}
\end{eqnarray}
The Riemann tensor encodes the local geometry of ${\mathfrak M}$,
and is related to the parallel transport of vectors on ${\mathfrak
M}$. In particular, if we define the covariant derivative by
\begin{eqnarray}
\nabla_{j}A_i = \frac{\partial A_i}{\partial\theta^j} -
\Gamma^k_{\,ij} A_k, \label{eq:3.14}
\end{eqnarray}
then the commutator of the covariant derivatives defines the Riemann
tensor:
\begin{eqnarray}
\nabla_{j}\nabla_{k}A_i-\nabla_{k}\nabla_{j}A_i = A_l R^l_{\ ijk}.
\label{eq:3.141}
\end{eqnarray}
The symmetry properties of the Riemann tensor can be derived by
lowering the index with the metric and writing $R_{ijkl}=G_{im}
R^m_{\ jkl}$. Specifically, this is given by
\begin{eqnarray}
\fl R_{ijkl} = \half \left( \frac{\partial^2{G_{il}}}
{\partial\theta^j\partial\theta^k}+ \frac{\partial^2{G_{jk}}}
{\partial\theta^i \partial\theta^l}- \frac{\partial^2{G_{ik}}}
{\partial\theta^j\partial\theta^l}- \frac{\partial^2{G_{jl}}}
{\partial\theta^i\partial\theta^k} \right) + G_{nm}\left(
\Gamma^n_{\ jk}\Gamma^m_{\ il}- \Gamma^n_{\ jl}\Gamma^m_{\ ik}
\right), \label{eq:1.301}
\end{eqnarray}
whence it follows that $R_{ijkl}=R_{klij}$. Along with the relation
$R^i_{\ jkl}=-R^i_{\ jlk}$ that follows from (\ref{eq:3.8}) we find
that $R_{ijkl}=-R_{ijlk}=-R_{jikl}$. Therefore, the components of
$R_{ijkl}$ vanish when $i=j$ or $k=l$. In the case of a
two-dimensional manifold ${\mathfrak M}$ the only nonvanishing
components of the Riemann tensor are given by
\begin{eqnarray}
R_{1212}=-R_{1221}=-R_{2112}=R_{2121}.
\end{eqnarray}
In other words, $R_{1212}$ is the sole independent component in two
dimensions.

Given a Riemann tensor $R^i_{\ jkl}$ we define the associated Ricci
tensor by the symmetric expression
\begin{eqnarray}
R_{jl} = R^k_{\ jkl}.  \label{eq:3.142}
\end{eqnarray}
Equivalently, we can write $R_{jl}=G^{ik}R_{ijkl}$. A further
contraction with the metric defines the scalar curvature:
\begin{eqnarray}
R = {G}^{jl}R_{jl}. \label{eq:3.15}
\end{eqnarray}
We call (\ref{eq:3.15}) the \textit{Ricci scalar curvature}. If the
Ricci tensor $R_{jl}$ is proportional to the metric tensor
${G}_{jl}$ then the manifold ${\mathfrak M}$ is called an
\textit{Einstein space}. This is because the metric of such a space
satisfies the vacuum Einstein equation
\begin{eqnarray}
R^i_{\,k}-\half \delta^i_{\,k}R=0, \label{eq:1.401}
\end{eqnarray}
where $R^i_{\,k}=G^{il}R_{lk}$. The significance of the Einstein
equation in statistics or statistical mechanics can be seen if we
relax the normalisation condition on the square-root density
function $\xi(x)$ and thus eliminate the physically irrelevant
degree of freedom associated with the norm $\|\xi(x)\|$.
Specifically, if $T(x)$ represents an observable function on the
phase space ${\boldsymbol{\Gamma}}$ then its expectation value in
the generic `state' $\xi(x)$ is
\begin{eqnarray}
\langle T(x)\rangle = \frac{\int_{\boldsymbol{\Gamma}} \xi(x) T(x)
\xi(x)\rd x}{\int_{\boldsymbol{\Gamma}}\xi(x)^2\rd x}.
\label{eq:1.41}
\end{eqnarray}
Evidently, the expectation so defined is invariant under the scale
transformation $\xi(x)\to\lambda\xi(x)$, where $\lambda$ is an
arbitrary nonzero number. Thus, all the relevant statistical
information is encoded in the direction of the vector $\xi(x)\in
{\mathcal H}$, irrespective of its length. Via the identification
$\xi(x)\sim \lambda\xi(x)$ we obtain a space of rays in ${\mathcal
H}$, which is known as the real projective Hilbert space. Suppose
now that we consider the Einstein equation (\ref{eq:1.401}) for the
metric on the projective Hilbert space. Then there is a unique
solution which is induced by the infinitesimal form of the
Bhattacharyya spherical distance (\ref{eq:1.7}). Specifically, this
is obtained by setting $\xi_1(x)=\xi(x)$ and
$\xi_2(x)=\xi(x)+\rd\xi(x)$ in
\begin{eqnarray}
\cos^2\phi = \frac{\left(\int_{\mathds R} \xi_1(x) \xi_2(x)\rd x
\right)^2}{\left(\int_{\mathds R} \xi_1(x) \xi_1(x)\rd x\right)
\left(\int_{\mathds R} \xi_2(x) \xi_2(x)\rd x\right)},
\label{eq:1.10.1}
\end{eqnarray}
Taylor expanding each side, and retaining terms of quadratic order.
Then with the notation of (\ref{eq:1.11}) we can write
\begin{eqnarray}
\phi^2 = \frac{\langle\rd\xi,\rd\xi\rangle}{\langle\xi,\xi\rangle} -
\frac{\langle\xi,\rd\xi\rangle^2}{\langle\xi,\xi\rangle^2},
\label{eq:1.47}
\end{eqnarray}
which defines a metric on the projective Hilbert space. It follows
that the Einstein equation uniquely determines the probabilistic
properties of the space of densities.

For a statistical manifold ${\mathfrak M}$ associated with a
distribution of the exponential type (\ref{eq:3.1}), the Riemann
tensor assumes a simple form because the first two terms in
(\ref{eq:3.8}) cancel, and only the contractions of the Christoffel
symbols remain.

The examples of statistical mechanical systems considered here are
parameterised by a pair of external variables, so that the
statistical manifold ${\mathfrak M}$ is two-dimensional. In this
case, the expression for the scalar curvature admits further
simplifications. Specifically, we have the following:

\begin{prop}
In terms of the canonical parametrisation $(\theta^1,\theta^2)$, the
scalar curvature of a two-dimensional statistical model
corresponding to the density function {\rm(\ref{eq:3.1})} is given
by the determinant
\begin{eqnarray}
R = -\frac{1}{2{G}^2} \left| \begin{array}{lll}
\psi_{11}(\theta) & \psi_{12}(\theta) & \psi_{,22}(\theta) \\
\psi_{111}(\theta) & \psi_{112}(\theta) & \psi_{122}(\theta)\\
\psi_{112}(\theta) & \psi_{122}(\theta) & \psi_{222}(\theta),
\end{array} \right| , \label{eq:3.16}
\end{eqnarray}
where ${G}=\det({G}_{ij})$ is the determinant of the Fisher-Rao
metric, and where $\psi_{12}(\theta)=\partial^2
\psi(\theta)/\partial\theta^1\partial\theta^2$,
$\psi_{112}(\theta)=\partial^3\psi(\theta)/\partial\theta^1
\partial\theta^1\partial\theta^2$, and so on. \label{prop:2}
\end{prop}

{\it Proof}. Recall that the scalar curvature is defined by the
contraction
\begin{eqnarray}
R_{ijkl} = \left( \Gamma_{kmi}\Gamma_{jln}- \Gamma_{kmj}
\Gamma_{iln}\right) {G}^{mn}. \label{eq:3.17}
\end{eqnarray}
Substituting (\ref{eq:1.24}) and the inverse of (\ref{eq:1.23}) in
(\ref{eq:3.17}), we obtain (\ref{eq:3.16}) after some rearrangement
of terms. \hspace*{\fill} $\square$

\subsection{Geometry of Gaussian distributions}
\label{sec:1.4}

As an elementary illustrative example, consider the Gaussian
(normal) distribution $N(\mu,\sigma)$ on the real line ${\mathds R}$
with mean $\mu$ and variance $\sigma>0$. For the parameterised
density function we have
\begin{eqnarray}
p(x|\mu,\sigma) = \frac{1}{\sqrt{2\pi}\sigma}\exp\left(
-\frac{(x-\mu)^{2}}{2\sigma^{2}} \right) . \label{eq:3.24}
\end{eqnarray}
The normal density function can be rewritten in the canonical form
\begin{eqnarray}
p(x|\theta)=\exp\left[-\theta^1 x^2 -\theta^2 x
-\psi(\theta)\right], \label{eq:3.25}
\end{eqnarray}
where
\begin{eqnarray}
\theta^1=\frac{1}{2\sigma}, \quad \theta^2 = -\frac{\mu}{\sigma^2},
\quad {\rm and}\quad \psi(\theta)= \octa \left(
\frac{\theta^2}{\theta^1}\right)^2 -\ln\Big(
\frac{\sqrt{2\pi}}{2\theta^1} \Big). \label{eq:3.26}
\end{eqnarray}
By differentiating $\psi(\theta)$ with respect to the parameters
$\{\theta^i\}$ we can determine the components of the metric tensor
${G}_{ij}(\theta)$ in the coordinate system specified by the
canonical parametrisation $\{\theta^i\}$.

Alternatively, we may regard the mean $\mu$ and variance $\sigma$ as
coordinates on the statistical manifold ${\mathfrak M}$. In terms of
the parameters $(\mu,\sigma)$ the metric does not admit a simple
representation (\ref{eq:1.23}), and we must perform the Gaussian
integration in the defining relation (\ref{eq:1.13}). The line
element then becomes
\begin{eqnarray}
{\rm d}s^{2} = \frac{1}{\sigma^{2}}({\rm d}\mu^{2} + 2 {\rm
d}\sigma^{2}) \label{eq:3.27}
\end{eqnarray}
which is defined on the upper-half plane $-\infty<\mu<\infty$ and
$0<\sigma<\infty$. Since the metric $G_{ij}$ is diagonal in these
coordinates, it is easily inverted and we obtain
\begin{eqnarray}
{G}^{ij} = \sigma^2 \left( \begin{array}{cc} 1 & 0 \\ 0 & \half
\end{array} \right). \label{eq:3.28}
\end{eqnarray}
A short calculation shows that the Christoffel symbols are given by
\begin{eqnarray}
\Gamma^1_{12}=\Gamma^1_{21}=-2\Gamma^2_{11}=\Gamma^2_{22}=-
\frac{1}{\sigma}, \qquad \Gamma^1_{11}=\Gamma^1_{22}=
\Gamma^2_{12}=\Gamma^2_{21}=0 .  \label{eq:3.29}
\end{eqnarray}
Since the inverse of the metric tensor (\ref{eq:3.28}) is diagonal,
we need only determine the diagonal components of the Ricci tensor
in order to calculate the scalar curvature (that is, the
off-diagonal elements of the Ricci tensor vanish). These are
\begin{eqnarray}
R_{11}=-\frac{1}{2\sigma^2} \quad {\rm and} \quad R_{22}=-
\frac{1}{\sigma^2},
\end{eqnarray}
respectively. Hence, the resulting geometry is that of a hyperbolic
space, which is a homogeneous manifold of constant negative
curvature:
\begin{eqnarray}
R=-1. \label{eq:1.57}
\end{eqnarray}

This space has many interesting properties. For example, the
geodesic equations (\ref{eq:1.31}) characterising trajectories of
shortest paths on ${\mathfrak M}$ are determined by the equations
\begin{eqnarray}
\frac{\rd^2 \mu(s)}{\rd s^2} - 2 \frac{1}{\sigma(s)} \frac{\rd
\mu(s)}{\rd s}\frac{\rd \sigma(s)}{\rd s} = 0  \label{eq:3.30}
\end{eqnarray}
and
\begin{eqnarray}
\frac{\rd^2 \sigma(s)}{\rd s^2} + \frac{1}{\sigma(s)}\left\{
\frac{1}{2} \left(\frac{\rd \mu(s)}{\rd s}\right)^2-\left( \frac{\rd
\sigma(s)} {\rd s}\right)^2\right\} = 0. \label{eq:3.31}
\end{eqnarray}
Since $\sigma>0$ we can divide (\ref{eq:3.30}) and (\ref{eq:3.31})
by $\sigma$, obtaining
\begin{eqnarray}
\Big(\frac{\mu'}{\sigma}\Big)^\prime - \frac{\mu'}{\sigma}
\frac{\sigma'}{\sigma} = 0 \quad {\rm and} \quad \Big(
\frac{\sigma'}{\sigma}\Big)^\prime + \frac{1}{2}
\frac{\mu'}{\sigma}\frac{\mu'}{\sigma}=0, \label{eq:3.32}
\end{eqnarray}
where the prime indicates $\rd/\rd s$. It follows that
\begin{eqnarray}
\left( \Big(\frac{\mu'}{\sigma}\Big)^2 + 2 \Big(
\frac{\sigma'}{\sigma} \Big)^2 \right)^\prime = 0, \label{eq:3.33}
\end{eqnarray}
and hence that
\begin{eqnarray}
\Big(\frac{\mu'}{\sigma}\Big)^2 + 2 \Big( \frac{\sigma'}{\sigma}
\Big)^2 = v^2, \label{eq:3.34}
\end{eqnarray}
where $v\geq0$ is a constant. On the other hand, if we define
$X=\mu'/\sigma$ then from the first equation of (\ref{eq:3.32}) we
have $X\sigma'-X'\sigma=0$, or equivalently $(X/\sigma)'=0$, and
thus $\mu'/\sigma=c\,\sigma$, where $c$ is a constant. Substituting
this in (\ref{eq:3.34}) we deduce that
\begin{eqnarray}
c^2 \sigma^2 + \Big(\frac{\sigma'}{\sigma}\Big)^2=v^2,
\label{eq:3.35}
\end{eqnarray}
where we have rescaled the integration constants, i.e. $c
\to\sqrt{2}c$ and $v\to\sqrt{2}v$. There are now two cases to
consider, depending on whether $c$ is zero or nonzero.

If $c=0$, then $\mu$ is constant, and $\sigma'=v\sigma$, so that
$\sigma(s)=a\re^{vs}$ for some constants $a$ and $v$ such that $a>0$
and $v>0$. This represents a straight line parallel to the $\sigma$
axis in the $\mu$-$\sigma$ plane. If $c\neq0$, then $c\,\sigma\leq
v$ from (\ref{eq:3.35}), hence $\sigma(s)=\frac{v}{c}\sin\gamma(s)$
for some $\gamma(s)$. Substituting this into (\ref{eq:3.35}) we
obtain
\begin{eqnarray}
\left( \frac{\rd\gamma}{\rd s}\right)^2 = v^2 \sin^2\gamma(s).
\label{eq:3.36}
\end{eqnarray}
Since $\gamma^\prime\neq0$, $\gamma(s)$ is monotonic and thus
invertible. We may assume, without loss of generality, that
$\gamma^\prime>0$ so that (\ref{eq:3.36}) implies $\rd\gamma/\rd
s=v\sin\gamma$. Using $\sigma(s)=\frac{v}{c}\sin\gamma(s)$ we find
that
\begin{eqnarray}
\frac{\sigma'}{\sigma} = \frac{1}{\sigma}\frac{\rd\sigma}
{\rd\gamma}\frac{\rd\gamma}{\rd s}=v\cos\gamma, \label{eq:3.37}
\end{eqnarray}
and further, using (\ref{eq:3.34}) with rescaled $c$ and $v$ we
obtain
\begin{eqnarray}
\mu'(s)= v\sigma(s) \sin\gamma(s).
\end{eqnarray}
If we regard $s=s(\gamma)$ as parameterised by $\gamma$ we can write
\begin{eqnarray}
\mu(s)= \int \frac{\rd \mu}{\rd\gamma}\,\rd\gamma = \int \mu'
\frac{\rd s}{\rd\gamma}\,\rd\gamma = \frac{v}{c}\int \sin\gamma
\rd\gamma = -\frac{v}{c}\cos\gamma+b, \label{eq:3.38}
\end{eqnarray}
where $b$ is an integration constant.

\begin{figure}
\begin{center}
  \includegraphics[scale=0.8]{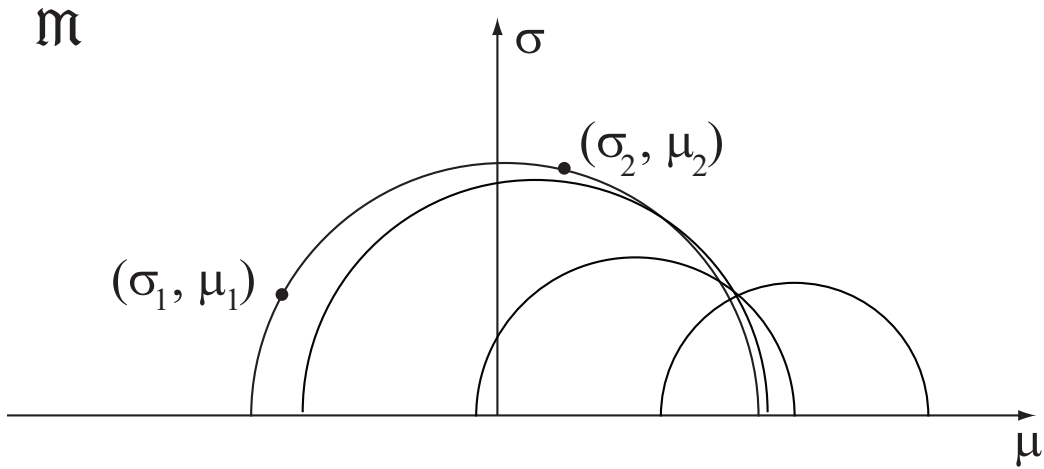}
  \caption{\label{fig:3.1}{\it Geodesic curves for Gaussian
 distributions}. The statistical manifold ${\mathfrak M}$ in this
 case is the upper half plane parameterised by $\mu$ and $\sigma$.
 We have $-\infty<\mu<\infty$ and $0<\sigma<\infty$. The shortest
 path joining the two normal distributions $N(\mu_{1},\sigma_{1})$
 and $N(\mu_{2},\sigma_{2})$ is given by the unique semi-circular
 arc through the given two points and centred on the boundary line
 $\sigma=0$.
  }
\end{center}
\end{figure}

To summarise, if we regard $\gamma(s)$ as the independent parameter,
then the solutions to the geodesic equations for the Gaussian family
of densities (\ref{eq:3.24}) are
\begin{eqnarray}
\mu(s)=-\frac{v}{c}\cos\gamma(s)+b \quad {\rm and} \quad \sigma(s)=
\frac{v}{c} \sin\gamma(s). \label{eq:3.39}
\end{eqnarray}
These equations represent half-circles on the $\mu$-$\sigma$ plane
centred on the $\mu$ axis ($\sigma=0$) with radius $v/c$. In
Figure~\ref{fig:3.1} we sketch examples of geodesic curves for the
Gaussian family, for $c\neq0$.

If follows from the solutions to the geodesic equations that the
separation of a pair of normal distributions $N(\mu_{1},\sigma_{1})$
and $N(\mu_{2},\sigma_{2})$ is given by
\begin{eqnarray}
D(\rho_{1},\rho_{2}) = \frac{1}{\sqrt{2}}\log
\frac{1+\delta_{1,2}}{1-\delta_{1,2}} , \label{eq:1.69}
\end{eqnarray}
where the function $\delta_{1,2}$ defined by
\begin{eqnarray}
\delta_{1,2}=\sqrt{\frac{(\mu_2-\mu_1)^2+2(\sigma_2 -\sigma_1)^2}
{(\mu_2-\mu_1)^2+2(\sigma_2+\sigma_1)^2}} \label{eq:3.41}
\end{eqnarray}
lies between 0 and 1. These results follow directly from the fact
that the geodesics are semi-circular arcs centred on the boundary
line $\sigma=0$ (this line itself is not part of the manifold
${\mathfrak M}$ because $\sigma>0$). In the exceptional case when
$\mu_{1}= \mu_{2}$, the geodesic is a straight line $\mu$ =
constant, and
\begin{eqnarray}
D(\rho_{1},\rho_{2}) = \frac{1}{\sqrt{2}}\left|\log
\frac{\sigma_{1}}{\sigma_{2}}\right|. \label{eq:1.71}
\end{eqnarray}

The above example illustrates how various geometric aspects of a
statistical manifold ${\mathfrak M}$ can be investigated in a
systematic manner. It is interesting to note, in particular, that
the Gaussian distributions define an elementary hyperbolic geometry
with constant negative curvature. Before examining various geometric
characterisations of ideal and interacting gases in thermal
equilibrium, let us discuss the relation between the Fisher-Rao
distance measure and various measures of entropy, a topic of some
interest.

\subsection{From entropy to Fisher information}
\label{sec:1.5}

We have observed how the notion of information geometry arises from
the Fisher information matrix commonly used in statistical analysis.
On the other hand, the term `information' often suggests the concept
of entropy, rather than the Fisher matrix. Indeed, the entropy
concept is essential in thermodynamics and statistical mechanics.
Therefore, it would be appropriate to clarify the interrelation
between these two notions of information.

We shall discuss entropy in a fairly general context, and consider
as before a parametric family of probability density functions
$p(x|\theta)$ which we assume to be defined, say, on the real line
${\mathds R}$, where $\theta=\{ \theta^i\}$. Then with respect to
any twice-differentiable concave function $\varphi(p)$ we define the
associated \textit{entropy functional} by the expression
\begin{eqnarray}
H_\varphi(p) = \int_{\mathds R} \varphi[p(x|\theta)]\rd x.
\end{eqnarray}
Now, if $f$ represents a vector in the tangent space (at $p$) of the
manifold of density functions, then the derivative of the entropy
$H_\varphi$ at $p$ in the direction $f$ is defined by
\begin{eqnarray}
\rd H_\varphi(p;f) &=& \left. \frac{\rd}{\rd s}\, H_\varphi(p+sf)
\right|_{s=0} \nonumber\\ &=&\int_{\mathds R}\varphi^{\prime}[p(x)]
f(x)\rd x,
\end{eqnarray}
where $\varphi^{\prime}(p)=\rd\varphi(p)/\rd p$. Similarly, if $f$
and $g$ are two vectors in the tangent space at $p$, we define the
Hessian of $H_\varphi$ by
\begin{eqnarray}
\rd^2 H_\varphi(p;f,g) = \int_{\mathds R}\varphi^{\prime\prime}
[p(x)] f(x)g(x) \rd x.
\end{eqnarray}
The corresponding quadratic form is
\begin{eqnarray}
\triangle_f H_\varphi(p)=4\rd^2H_\varphi(p;f,f),
\end{eqnarray}
or equivalently,
\begin{eqnarray}
\triangle_fH_\varphi(p) = 4\int_{\mathds R} \varphi^{\prime\prime}
[p(x)] f^2(x)\rd x,
\end{eqnarray}
where the factor of $4$ here is purely conventional. The concavity
of $\varphi$ then implies that
\begin{eqnarray}
-\triangle_fH_\varphi(p)\geq0.
\end{eqnarray}
In particular, if we chose $f$ to be $\partial_ip$, where
$\partial_i =\partial/\partial\theta^i$, then we have
\begin{eqnarray}
\triangle_{\partial p}H_\varphi(p) = 4\int_{\mathds R}
\varphi^{\prime\prime} [p(x|\theta)]\left(\partial_i p(x|\theta)
\right)^2 \rd x. \label{eq:75}
\end{eqnarray}

Thus far, we have not specified the form of the function $\varphi$,
except for the requirements of concavity and twice
differentiability. As a special case, let us consider the
one-parameter family of concave functions
\begin{eqnarray}
\varphi_\alpha(z)=(\alpha-1)^{-1}(z-z^\alpha), \label{eq:xx1}
\end{eqnarray}
where $\alpha>0$. This determines a one-parameter family of
entropies $H_\alpha(p)$ given by
\begin{eqnarray}
H_\alpha(p) = \frac{1}{\alpha-1}\left( 1-\int p^\alpha(x)\rd x
\right). \label{eq:1.80}
\end{eqnarray}
Note that when $\alpha=1$ we have $\varphi_1(z)=-z\ln z$ and hence
\begin{eqnarray}
H_1(p)=-\int p(x)\ln p(x)\rd x.
\end{eqnarray}
In other words, we recover the expression for the familiar
Shannon-Wiener entropy in the limit $\alpha\to1$. In the general
case, the expression (\ref{eq:1.80}) defines the
\textit{Havrda-Charv\'at entropy} (also known as the $\alpha$-order
entropy), which is related to the well-known \textit{R\'enyi
entropy} $R_\alpha(p)$ as follows:
\begin{eqnarray}
R_\alpha(p)=\frac{1}{1-\alpha}\ln\Big(1+(1-\alpha)H_\alpha(p) \Big).
\label{eq:2.007}
\end{eqnarray}
In particular, $R_\alpha$ is monotonic in $H_\alpha$.

Also, choosing $\varphi_\alpha$ as in (\ref{eq:xx1}) we find that
\begin{eqnarray}
\rd s_\alpha^2 = -\frac{1}{4\alpha}\,\triangle_{\partial{\vec p}}
H_\varphi(p) = -\frac{1}{\alpha}\, \rd^2
H_\varphi(p;\partial_ip,\partial_jp)\rd\theta^i \rd\theta^j
\label{eq:1.82}
\end{eqnarray}
is positive definite and defines a Riemannian metric. We can
summarise this as follows.

\begin{prop}[Burbea-Rao metric]
The coefficients of the differential metric
\begin{eqnarray}
\rd s_\alpha^2={G}_{ij}^{(\alpha)}\rd \theta^i \rd\theta^j
\end{eqnarray}
associated with the Hessian of the $\alpha$-order entropy {\rm
(\ref{eq:1.80})} are
\begin{eqnarray}
{G}_{ij}^{(\alpha)}=\int p^\alpha(\partial_i\ln p) (\partial_j\ln p)
\rd x. \label{eq:1.85}
\end{eqnarray}
In particular, when $\alpha=1$, ${G}_{ij}^{(\alpha)}$ reduces to the
Fisher-Rao metric. \label{prop:3}
\end{prop}

{\it Proof}. The expression in (\ref{eq:1.85}) follows at once from
(\ref{eq:1.82}) by virtue of the relation
$\varphi_\alpha^{\prime\prime}(p)=-\alpha p^{\alpha-2}$.
\hspace*{\fill} $\square$

We find, therefore, that the so-called $\alpha$-order entropy metric
is closely related to the Fisher-Rao geometry of the statistical
manifold. In addition, there is another significant relationship
between the derivative of the entropy and the $\alpha$-order Fisher
information matrix. This can be established as follows. From the
defining equation (\ref{eq:1.80}) we have
\begin{eqnarray}
\partial_i\partial_j H_\alpha(p)=-\alpha\int_{\mathbb
R}p^{\alpha-2} (\partial_i p)(\partial_j p)\rd x -
\frac{\alpha}{\alpha-1} \int_{\mathbb
R}p^{\alpha-1}\partial_i\partial_j p\,\rd x,
\end{eqnarray}
and therefore we deduce that
\begin{eqnarray}
{G}_{ij}^{(\alpha)}=-\frac{1}{\alpha}\partial_i\partial_j
H_\alpha(p) + \frac{1}{\alpha-1} \int_{\mathds R}p^{\alpha-1}
\partial_i \partial_j p\,\rd x.
\end{eqnarray}
For the canonical distribution (\ref{eq:3.1}), the limit
$\alpha\to1$ of this relation yields the Shannon-Wiener entropy
\begin{eqnarray}
H_1(p)=\sum_i\theta^i\int p(x|\theta)H_i(x)\rd x + \psi(\theta).
\end{eqnarray}
In other words, the thermodynamic potential and the entropy are
related by a Legendre transformation. Consequently, the Fisher-Rao
geometry and the geometry arising from the Hessian of the
Shannon-Wiener entropy are related by the general theory of Legendre
transforms.

\section{Classical ideal gas}
\label{sec:2}

We shall now characterise the geometry of the statistical manifold
that arises from the equilibrium distribution of a gas of
noninteracting particles. Although this system displays no phase
transition, the analysis presented here will provide an enlightening
contrast with the results of Section~\ref{sec:3} where we shall
examine the geometry of the van der Waals gas, which does exhibit a
liquid-vapour transition.

\subsection{Partition function in $P$-$T$ distribution}
\label{sec:2.1}

To elucidate the geometrical representation of gaseous systems in
statistical mechanics, we begin our analysis with a system of
noninteracting identical particles in the absence of potential
energy. Physically, this system corresponds to a classical ideal gas
immersed in a heat bath. As we shall show below, not only the
Riemann curvature but also the geodesic equations for this system
can be solved exactly. We consider, in particular, a
pressure-temperature ($P$-$T$) distribution (also known as the
Boguslavski distribution) of the form
\begin{eqnarray}
p(H,V|\alpha,\beta) = \frac{\exp\left(- \beta H - \alpha
V\right)}{Z(\alpha,\beta)} \label{eq:3.43}
\end{eqnarray}
defined on the phase space $\boldsymbol{\Gamma}$ of a system of
noninteracting particles. Here the partition function $Z(\alpha,
\beta)$ is determined by the phase-space and volume integral
\begin{eqnarray}
Z(\alpha,\beta) = \frac{1}{N!h^{3N}} \int_{0}^{\infty}\left(
\int_{\boldsymbol{\Gamma}} \exp\left(-\beta H\right){\rm dqdp}
\right) \exp\left(-\alpha V\right) {\rd}V . \label{eq:3.44}
\end{eqnarray}
As usual, we have $\beta=1/k_BT$, $\alpha=P/k_BT$, where $P$ denotes
the
pressure, $h$ the Planck constant, and $N$ the number of
particles. The Hamiltonian $H$ is just the free particle kinetic
energy
\begin{eqnarray}
H = \sum_{i=1}^{N}\frac{{\rm p}_{i}^{2}}{2m}. \label{eq:3.45}
\end{eqnarray}
Thus, we consider a closed system of noninteracting gas molecules
immersed in a heat bath at inverse temperature $\beta$ and effective
pressure $\alpha$. Since the system is in contact with a bath at
fixed temperature and pressure, the system energy and volume
fluctuate. In thermal equilibrium, the distribution of these
variables is determined by (\ref{eq:3.43}). For a real gas, the
constituent particles inevitably interact. Nevertheless, the ideal
gas represented by the distribution (\ref{eq:3.43}) adequately
characterises the properties of a real gas at high temperature or
low density, where the effects of inter-particle interactions can be
neglected.

Comparing (\ref{eq:3.43}) and (\ref{eq:3.1}) we observe that the
thermodynamic potential is given by $\psi(\alpha,\beta) = \ln
Z(\alpha,\beta)$. Therefore, to determine the Fisher-Rao metric
(\ref{eq:1.23}) we must perform the integration (\ref{eq:3.44}).
Noting the fact that each q-integration in (\ref{eq:3.44}) gives the
volume $V$ of the system, one obtains the partition function
\begin{eqnarray}
Z(\alpha,\beta) = \left( \frac{2\pi m}{h^{2}\beta} \right)^{3N/2}
\alpha^{-(N+1)}.  \label{eq:3.46}
\end{eqnarray}
This follows from the fact that
\begin{eqnarray}
\int_{\boldsymbol{\Gamma}} \exp\left(-\beta H\right){\rm dqdp} =
\int_{\rm q} \left(\int_{\rm p} \re^{-\beta\sum_{i=1}^{N}
\frac{{\rm p}_{i}^{2}}{2m}}\prod_{i=1}^N \rd{\rm p}_i\right)
\prod_{i=1}^N \rd{\rm q}_i \label{eq:3.47}
\end{eqnarray}
is just a product of Gaussian integrals, and the identity
\begin{eqnarray}
\frac{1}{N!} \int_{0}^\infty V^N \re^{-\alpha V} \rd V =
\alpha^{-(N+1)}  \label{eq:3.48}
\end{eqnarray}
that holds for $\alpha>0$.

Note that the partition function $z(\beta,V)$ in the canonical
ensemble is
\begin{eqnarray}
z(\beta,V) &=& \frac{1}{N!h^{3N}} \int_{\boldsymbol{\Gamma}}
\exp\left(-\beta H\right){\rm dqdp} \nonumber \\ &=& \frac{1}{N!}
\left(\frac{2\pi m}{h^{2}\beta}\right)^{3N/2}V^N, \label{eq:3.444}
\end{eqnarray}
from which one can calculate the \textit{Helmholtz free energy}
\begin{eqnarray}
F(\beta,V)=-k_BT\ln z(\beta,V)
\end{eqnarray}
and thus obtain the equation of
state
\begin{eqnarray}
P = -\left(\frac{\partial F}{\partial V}\right)_\beta = Nk_BT/V
\label{eq:3.445}
\end{eqnarray}
satisfied by a classical ideal gas.

\subsection{Curvature and geodesics for ideal gas}
\label{sec:2.2}

The expression (\ref{eq:3.46}) for the partition function clearly
shows that the Riemannian geometry of the statistical model
${\mathfrak M}$ associated with the classical ideal gas depends upon
the number $N$ of particles. Although finite size effects in small
systems are sometimes of interest, here we are primarily concerned
with the geometry that arises in the so-called thermodynamic limit
$N\to\infty$. Thus, we consider the thermodynamic potential
$\psi(\alpha,\beta)$ per particle in the thermodynamic limit, given
by
\begin{eqnarray}
\psi(\alpha,\beta) = \lim_{N\rightarrow\infty} N^{-1} \ln
Z(\alpha,\beta) = \frac{3}{2} \ln \frac{2\pi m}{h^{2}\beta} -
\ln\alpha .  \label{eq:3.49}
\end{eqnarray}
The components of the Fisher-Rao metric, with respect to the
parameterisation $(\alpha,\beta)$, can then be calculated by
differentiation, with the result
\begin{eqnarray}
{G}_{ij} = \left( \begin{array}{cc}
\alpha^{-2}  &  0  \\
0  &  \frac{3}{2}\beta^{-2}  \\
\end{array}  \right) . \label{eq:3.52}
\end{eqnarray}
From this expression we deduce the following.

\begin{prop}\label{prop:4}
All components of the Riemann tensor, and consequently also the
scalar curvature, of the statistical manifold ${\mathfrak M}$
associated with the classical ideal gas vanish, and thus the
manifold is flat.
\end{prop}

{\it Proof}. From the components of the metric (\ref{eq:3.52}) one
can calculate the components of the Christoffel symbol
$\Gamma^i_{jk}$ and the Riemann tensor $R^i_{\ jkl}$ using the
definitions (\ref{eq:1.24}) and (\ref{eq:3.8}). Alternatively, to
show that this manifold is flat, it suffices to display a change of
coordinates which transforms the metric (\ref{eq:3.52}) into a
Euclidean metric. Here, we adopt the latter approach because this
also permits more expeditious solution of the geodesic equations. We
recall that under a coordinate transformation $x^i\to{\bar x}^i$ the
metric of a Riemannian manifold transforms in the usual tensorial
manner, so that the components of the inverse metric in the new
coordinate system are determined by the contraction
\begin{eqnarray}
{\bar{G}}^{ij} = {G}^{kl}\,\frac{\rd{\bar x}^i}{\rd x^k}
\frac{\rd{\bar x}^j}{\rd x^l}.
\end{eqnarray}
Consider the following coordinate transformation
\begin{eqnarray}
\alpha\to\alpha' = \ln\alpha \quad {\rm and}\quad \beta\to \beta' =
\sqrt{{\textstyle\frac{3}{2}}}\ln\beta.
\end{eqnarray}
A straightforward calculation then shows that the components of
the inverse metric in the $(\alpha',\beta')$ coordinate system are
\begin{eqnarray}
\bar{G}^{ij} = \left( \begin{array}{cc}
1  &  0  \\
0  &  1  \\
\end{array}  \right) , \label{eq:3.53}
\end{eqnarray}
and thus the manifold is indeed flat. \hspace*{\fill} $\square$

Since the statistical manifold associated with the ideal gas is
flat, solution of the geodesic equations is straightforward. The
result can be summarised as follows.

\begin{prop}\label{prop:5}
The geodesic curves on the statistical manifold ${\mathfrak M}$
associated with the classical ideal gas are given by
\begin{eqnarray}
\frac{P}{P_0} = \left( \frac{k_BT}{k_BT_0}\right)^{1+c},
\label{eq:3.54}
\end{eqnarray}
where $P_0$, $T_0$, and $c$ are integration constants. In
particular, the geodesics include the adiabatic equation of state
for the ideal gas, corresponding to the choice $c=-C_V/Nk_B$, where
$C_V$ is the constant-volume heat capacity.
\end{prop}

{\it Proof}. The geodesic equations for the variables $\alpha$ and
$\beta$ assume identical forms, i.e.
\begin{eqnarray}
\frac{\rd^2x}{\rd s^2} - \frac{1}{x}\Big(\frac{\rd x}{\rd s}
\Big)^2=0 \label{eq:3.55}
\end{eqnarray}
for $x=\alpha,\beta$. This can be rewritten as
\begin{eqnarray}
\frac{\rd}{\rd s}\ln\Big(\frac{\rd x}{\rd s} \Big) =
\frac{\rd}{\rd s} \ln x, \label{eq:3.56}
\end{eqnarray}
from which we see that the general solution is $x(s)=c_1\re^{c_2s}$.
Thus, we obtain
\begin{eqnarray}
\frac{P}{k_BT} = c_1\,\re^{c_0s} \quad {\rm and}\quad
\frac{1}{k_BT} = c_3\,\re^{c_2s}  \label{eq:3.57}
\end{eqnarray}
as the general solution to the geodesic equations. Combining
these two equations, we have
\begin{eqnarray}
P = \frac{c_1}{c_3^{c_0/c_2}} \left(k_BT\right)^{1-c_0/c_2}.
\label{eq:3.58}
\end{eqnarray}
Setting $s=0$ we find $c_1=P_0/k_BT_0$ and $c_3=1/k_BT_0$, which
yields at once the expression in (\ref{eq:3.54}). \hspace*{\fill}
$\square$

\section{Van der Waals gas}
\label{sec:3}

The geometry of the statistical manifold changes considerably if the
gas molecules interact. In particular, if the system exhibits a
phase transition, then the curvature tends to become singular at the
transition point. This property seems to be universal and appears in
many systems exhibiting critical phenomena. The van der Waals gas
model is not only of physical interest, but also illustrates many of
the universal geometrical features of the associated manifold of
equilibrium states.

\subsection{Equation of state}
\label{sec:3.1}

\begin{figure}[t]
\begin{center}
  \includegraphics[scale=1.0]{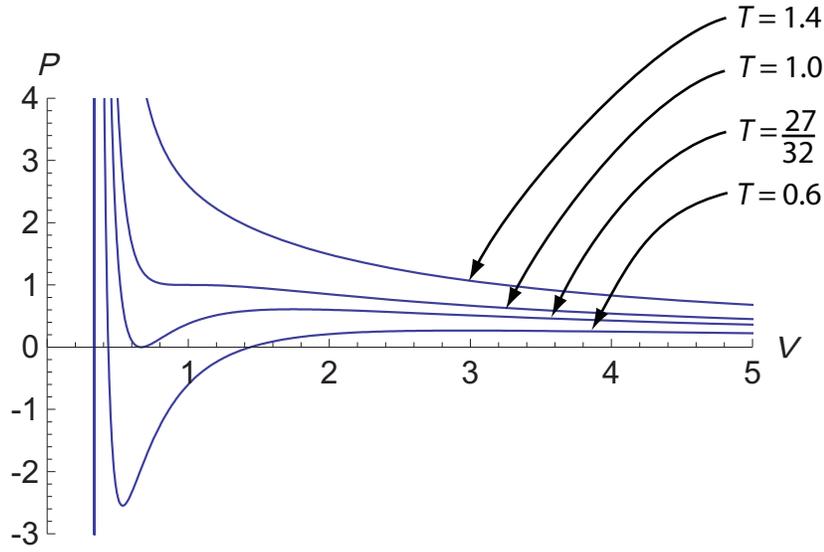}
  \caption{\label{fig:3-2}Equations of state for the van der Waals
 gas in terms of dimensionless variables. The isothermal curves
 correspond to $\hat{T}=1.4$,
 $\hat{T}=1$, $\hat{T}=27/32$ (maximum superheating temperature)
 and $\hat{T}=0.6$. Note that the  isothermal curves associated
 with temperatures below $T_{\rm m}$ allow metastable regions
 for which $\hat{P}<0$.
  }
\end{center}
\end{figure}

The idealised system of noninteracting particles considered above is
inadequate for the description of phase transition phenomena, that
is, the condensation of gas molecules. Here we shall extend the
model to include inter-particle interactions, which leads to the
\textit{van der Waals equation of state}:
\begin{eqnarray}
\left( P+a\frac{N^2}{V^2}\right) (V-b N)=Nk_BT , \label{eq:3.59}
\end{eqnarray}
where $N$ is the total number of molecules and $a,b$ are constants
determined by the properties of the molecule. The liquid-vapour
transition occurs at the critical point where the temperature $T$,
pressure $P$, and volume $V$ simultaneously assume
the values
\begin{eqnarray}
P_c=\frac{a}{27b^2}, \quad V_c=3b N, \quad {\rm and}\quad
T_c=\frac{8a}{27k_Bb}.
\end{eqnarray}
The critical point is determined by the simultaneous solution of the
equations
\begin{eqnarray}
\frac{\partial P}{\partial V}=0 \quad {\rm and} \quad
\frac{\partial^2 P}{\partial V^2}=0. \label{eq:3.63}
\end{eqnarray}
Using the dimensionless variables
\begin{eqnarray}
\hat{P}=\frac{P}{P_c}, \quad \hat{V}=\frac{V}{V_c}, \quad {\rm
and} \quad \hat{T}=\frac{T}{T_c},
\end{eqnarray}
the equation of state can be rewritten in the universal form
\begin{eqnarray}
\left(\hat{P}+3\hat{V}^{-2}\right)\left(\hat{V}-\third\right) =
{\textstyle\frac{8}{3}} \hat{T}, \label{eq:VDW}
\end{eqnarray}
independent of the parameters $a$ and $b$. In Figure~\ref{fig:3-2}
we plot the pressure $\hat{P}$ as a function of the volume
$\hat{V}$ for $\hat{T}>1$, $\hat{T}=1$, and $\hat{T}<1$.

Note that the positivity of the pressure implies a bound on
the temperature. In particular, from (\ref{eq:VDW}) we deduce that
the
condition $\hat{P}\geq0$ is equivalent to the bound:
\begin{eqnarray}
\hat{T}\geq\frac{9\hat{V}-3}{8\hat{V}^2},
\end{eqnarray}
in terms of the dimensionless volume ${\hat V}$. If we demand the
positivity of $\hat{P}$ for all volumes $\hat{V}\geq
\frac{1}{3}$, then we must require $\hat{T}\geq \frac{27}{32}$, or
equivalently
\begin{eqnarray}
T\geq \frac{27}{32}T_c.
\end{eqnarray}
The temperature
\begin{eqnarray}
T_{\rm m}=\frac{27}{32}T_c\approx0.85T_c
\end{eqnarray}
is known as the \textit{temperature of maximum superheating} and is
related to the nucleation of bubbles when the liquid is heated
very abruptly. In particular, if $T<T_{\rm m}$ then the liquid can
be contained at low external pressure, whereas if $T>T_{\rm m}$ the
liquid cannot exist under low external pressure and thus evaporates.
Experimental data show, for example, that $T_{\rm m}=0.89T_c$ for
ether, $T_{\rm m}=0.92T_c$ for alcohol, and $T_{\rm m}=0.84T_c$ for
water, indicating the fairly accurate predictive power of the van
der Waals equations of state.

\begin{figure}[t]
\begin{center}
  \includegraphics[scale=0.7]{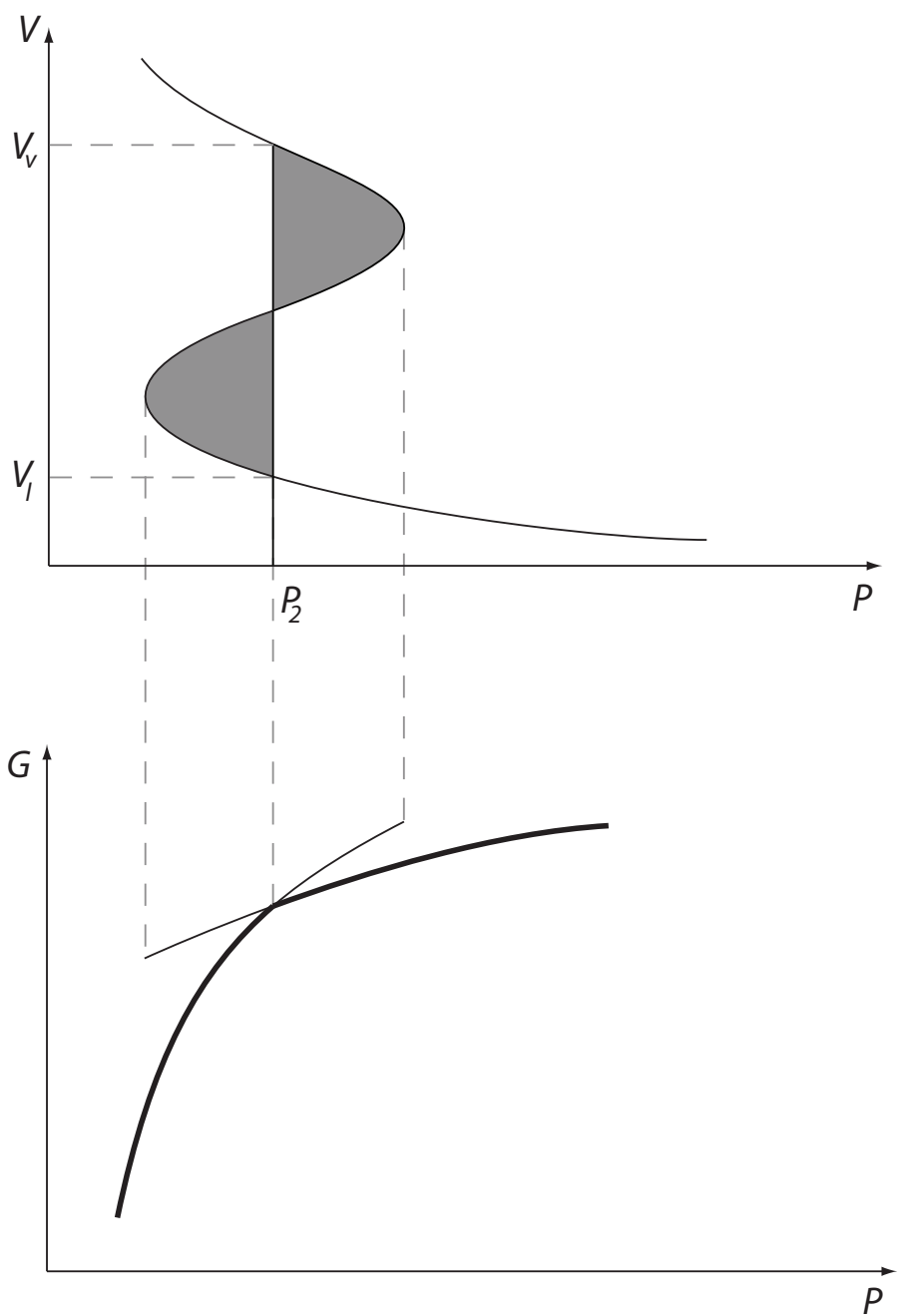}
  \caption{\label{fig:3.4} Isothermal curve and equal area law in
  the pressure-volume plane (above), and pressure dependence
 of the Gibbs free energy (below). As the pressure $P$ of the gas
 is slowly increased, the Gibbs free energy $G(P)$ increases along
 the thick solid line in the lower diagram, until $P$ reaches the
 coexisting pressure $P_2$. The gas then undergoes a phase
 transition and condenses. During this transition the volume
 changes from $V_v$ to $V_l$, at which point the entire system
 enters the liquid phase. The value of the coexisting
 pressure $P_2$ is determined by Maxwell's equal area law.
  }
\end{center}
\end{figure}

Turning to the equations of state, the pressure $P$ as a function of
the volume $V$ has three distinct roots when the temperature is
below its critical value $T_c$. Amongst these three roots, the
intermediate root corresponds to a point at which $(\partial
P/\partial V)_T>0$. Hence, this root is unstable, since the pressure
increases with volume for fixed temperature. It follows that one of
the remaining two roots should correspond to thermal equilibrium. To
ascertain which of the two roots is stable, we recall that the
condition for stability is determined by the minimisation of the
free energy. If we let $G(T,P)$ denote the Gibbs free energy, then
Maxwell's relation $V=(\partial G/\partial P)_T$ implies that
\begin{eqnarray}
G(T,P) = G(T,P_0) + \int_{P_0}^P V(u,T)\rd u.
\end{eqnarray}
Therefore, when viewed as a function of pressure $P$ for a fixed
temperature $T<T_c$ below the critical point, the free energy
$G(T,P)$ describes one of two distinct curves, depending on whether
$P$ is reduced from high values or increased from low values. This
is shown schematically in Figure~\ref{fig:3.4}.

If the free energy assumes its minimum value, then as the value of
$P$ changes, $G(T,P)$ must describe one of the two curves in Figure
\ref{fig:3.4} until its intersection with the other curve at
pressure $P=P_2$, whereafter $G(T,P)$ follows the other curve. At
the point $P=P_2$ the liquid and vapour phases coexist. Therefore,
if we, say, reduce the pressure from high values, then after
reaching the value $P_2$ the pressure remains constant until the
liquid is entirely converted into vapour. During this transition the
volume changes from $V_l$ to $V_v$ as indicated in
Figure~\ref{fig:3.4}. The value of the coexisting pressure $P_2$ is
determined, for each fixed $T<T_c$, by Maxwell's equal area
principle. That is, the vertical line in Figure \ref{fig:3.4} is
chosen so that the volumes of the two shaded regions are exactly
equal.

\begin{figure}[t]
\begin{center}
  \includegraphics[scale=1.0]{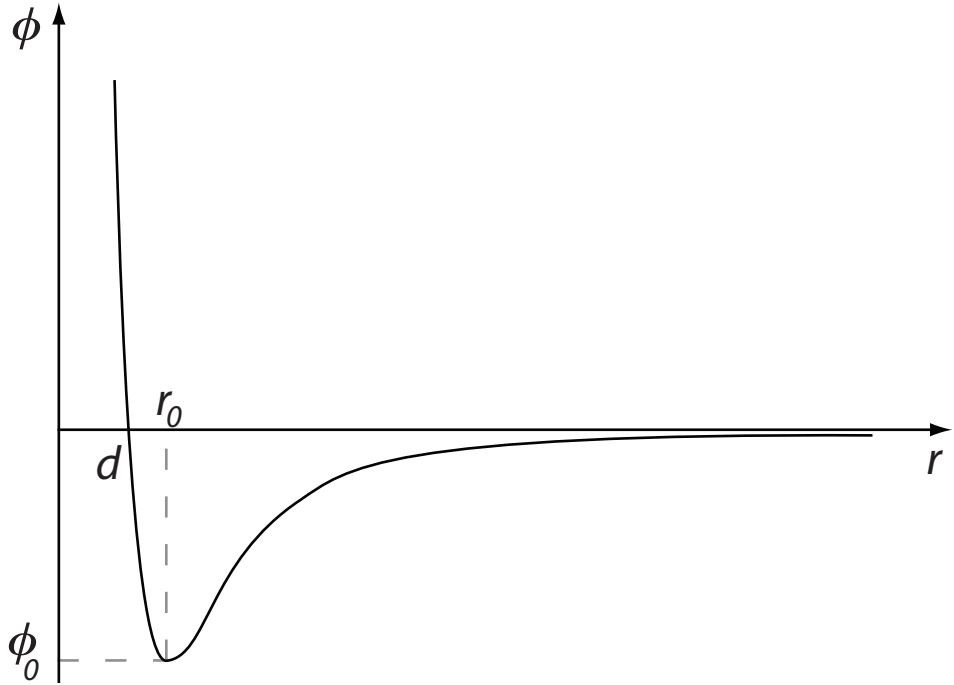}
  \caption{\label{fig:1.4}Lennard-Jones Potential. There is a strong
 repulsive force at short distances $r\leq r_0$ and a weak attractive
 force at long distances $r>r_0$, where $r_0=\sqrt[6]{2}d$ and $d$
 represents the radius of the gas molecule.
 }
\end{center}
\end{figure}

\subsection{Canonical partition function}
\label{sec:3.2}

The equation of state (\ref{eq:3.59}) was first deduced empirically
by van der Waals, directly from experimental observations. However,
it can also be derived analytically from the canonical partition
function associated with an empirically postulated intermolecular
potential. Assume that the interaction energy between a pair of
molecules separated by a distance $r$ is given by the Lennard-Jones
potential
\begin{eqnarray}
\phi(r) = 4\phi_0\left[ \Big( \frac{d}{r}\Big)^{12} - \Big(
\frac{d}{r} \Big)^6\right] = \phi_0 \left( \frac{r_0}{r}
\right)^{12} - 2\phi_0\left( \frac{r_0}{r}\right)^6 ,
\label{eq:L-J}
\end{eqnarray}
where $r_0=2^{1/6}d$ and $d$ is a parameter which can be regarded as
the radius of the gas molecule. Clearly, $\phi(d)=0$ and $\phi(r)$
assumes its minimum value at $r=r_0$. As we see from
Figure~\ref{fig:1.4}, this inter-molecular potential energy gives
rise to a weak long-range attractive force and a strong short-range
repulsive force between each pair of molecules.

The canonical partition function can thus be written as
\begin{eqnarray}
z(\beta,V) = \frac{1}{N!h^{3N}} \prod_{i=1}^N \int \rd^3{\rm p}_i
\int\rd^3r_i \exp\Big( -\beta \sum_{i=1}^N \frac{{\rm p}_i^2}{2m} -
\beta \sum_{(ij)} \phi_{ij}\Big),
\end{eqnarray}
where $\phi_{ij}=\phi(r_{ij})$, with $r_{ij}$ denoting the distance
between the $i$th and $j$th molecules. Thus, the canonical partition
function can be expressed as a product
\begin{eqnarray}
z(\beta, V)=z_0(\beta,V) Q(\beta,V),
\end{eqnarray}
where $z_0(\beta,V)$ is the canonical partition function for the
ideal gas (\ref{eq:3.444}) and
\begin{eqnarray}
Q(\beta,V) = \frac{1}{V^N} \int\rd^3r_1\cdots\int\rd^3r_N \exp
\Big(-\beta \sum_{(ij)} \phi_{ij}\Big) \label{eq:3.777}
\end{eqnarray}
is the contribution from the interaction energy.

Now, as an approximation to the Lennard-Jones potential we assume
that $\exp(-\beta\phi_{ij})=0$ for $r_{ij}<d$. In other words, we
regard the molecules as hard spheres of radius $d$, which cannot
overlap. As a consequence, the overlapping region can be removed
from the range of the volume integration (\ref{eq:3.777}). Defining
the so-called \textit{Mayer function} $f_{ij}=f(r_{ij})$ by
\begin{eqnarray}
f_{ij} = \exp(-\beta\phi_{ij})-1,
\end{eqnarray}
we rewrite the integral (\ref{eq:3.777}) as
\begin{eqnarray}
\hspace{-1cm}
Q(\beta,V)&=&\frac{1}{V^N}\int_{r_1>d}\!\!\!\!\rd^3r_1\cdots
\int_{r_N>d}\!\!\!\!\rd^3r_N \prod_{(ij)} \left(1+f_{ij}\right)
\nonumber \\ &=& \frac{1}{V^N} \int_{r_1>d}\!\!\!\!\! \rd^3r_1\cdots
\int_{r_N>d}\!\!\!\!\rd^3r_N \Big(1 + \sum_{(ij)} f_{ij} +
\sum_{(ij)} \sum_{(kl)}f_{ij}f_{kl} + \cdots\Big). \label{eq:3.778}
\end{eqnarray}
Assuming that the parameter $\phi_0$ in (\ref{eq:L-J}) is
sufficiently small, the contribution arising from $f_{ij}$ in the
specified integration range can be regarded as an infinitesimal. The
first term on the right side of (\ref{eq:3.778}), i.e. the integral
of unity, can, on the other hand, be approximated by
\begin{eqnarray}
V(V-v_0)\cdots[V-(N-2)v_0][V-(N-1)v_0] \approx \left( V-bN
\right)^N,
\end{eqnarray}
where we put $v_0=2b=\frac{4}{3}\pi d^3$. The integrations are
performed consecutively, so that the first particle can occupy
volume
$V$ without constraints, the second particle can occupy volume $V$
less the volume $v_0$ occupied by the first particle, the third
particle can occupy volume $V$ less the volume $2v_0$ occupied by
the first two particles, and so on. Similarly, the second term on
the right side of (\ref{eq:3.778}) can be approximated by
\begin{eqnarray}
\int_{r_1>d}\!\!\!\!\! \rd^3r_1\cdots\int_{r_N>d}\!\!\!\!\rd^3r_N
f_{ij} &=& \left( V-bN \right)^{N-1}\int_{r_j>d}\!\!\!\!\!
\rd^3r_j f_{ij} \nonumber \\ &\approx& -\left( V-bN \right)^{N-1}
\beta\pi \int_d^\infty \phi(r) r^2\rd r.
\end{eqnarray}

Assembling these results, we can approximate $Q(\beta,V)$
in the following closed form
\begin{eqnarray}
Q(\beta,V) &\cong& \left( 1-b\frac{N}{V}\right)^{N}\left( 1+
\beta\frac{aN^2}{V}+\cdots\right) \nonumber \\ &\cong& \left(
1-b\frac{N}{V}\right)^{N}\exp\left(\beta\frac{aN^2}{V}\right),
\end{eqnarray}
where we have defined
\begin{eqnarray}
a = -\pi\int_d^\infty r^2 \phi(r)\rd r.
\end{eqnarray}
Using the above expression for $Q(\beta,V)$ we finally obtain
the canonical partition function
\begin{eqnarray}
z(\beta,V) = \frac{1}{N!} \left( \frac{2\pi m}{\beta h^{2}}
\right) ^{3N/2} (V-bN)^{N} \exp(a\beta N^{2}/V). \label{eq:3.76}
\end{eqnarray}

\subsection{Critical behaviour of the van der Waals gas}
\label{sec:3.3}

From the expression for the partition function of the canonical
distribution we deduce the equation of state
\begin{eqnarray}
P=\frac{1}{\beta}\left(\frac{\partial\ln z(\beta,V)}{\partial V}
\right)_\beta =\frac{Nk_BT}{V-bN}-a\frac{N^2}{V^2},
\label{eq:3.77}
\end{eqnarray}
where for clarity we have substituted $\beta=1/k_BT$. Observe that
this is precisely the van der Waals equation introduced in
(\ref{eq:3.59}). If we had not applied various approximations in the
derivation of (\ref{eq:3.76}), then additional terms of order
$(N/V)^3$ and higher would have appeared on the right side of
(\ref{eq:3.77}).

To analyse the behaviour of the system near a critical
point we introduce the deviation parameters
\begin{eqnarray}
p=\hat{P}-1, \quad v=\hat{V}-1, \quad {\rm and}\quad t=\hat{T}-1.
\end{eqnarray}
In terms of these shifted variables the equation of state
(\ref{eq:3.77}) becomes
\begin{eqnarray}
p=\frac{8(t+1)}{3v+2} - \frac{3}{(v+1)^2} - 1. \label{eq:3.130}
\end{eqnarray}
We can then expand the equation of state (\ref{eq:3.130}) for
small $v$ and $t$, obtaining
\begin{eqnarray}
p = t(4-6v) - {\textstyle\frac{3}{2}}v^3 + \cdots.\label{eq:vdw2}
\end{eqnarray}
Similarly the Gibbs free energy
\begin{eqnarray}
G = PV-k_BT\ln z(T,V) \label{eq:3.80}
\end{eqnarray}
can be expanded as follows:
\begin{eqnarray}
\fl G = P_cV_c\Big[ (p-4t)v+3tv^2+{\textstyle\frac{3}{8}}v^4+1+p
\Big] - {\textstyle\frac{3}{2}} Nk_BT\ln \left(\frac{2\pi mk_BT}
{h^2} \right). \label{eq:gibbs}
\end{eqnarray}
For fixed pressure $p$ and temperature $t$, the volume $v$ in
thermal equilibrium is that which minimises the Gibbs free energy
$G$. The equation of state (\ref{eq:vdw2}) is a necessary but not
sufficient condition for the Gibbs free energy to assume its
minimum. Therefore, at the coexisting pressure $p=4t$ below the
critical temperature $(t<0)$ we have the three roots
\begin{eqnarray}
v = \pm2\sqrt{-t}, \quad 0
\end{eqnarray}
for the volume determined by the equation of state (\ref{eq:vdw2}).
Differentiating (\ref{eq:gibbs}) with respect to $v$, we find that
the first two roots $\pm2\sqrt{-t}$ minimise the free energy $G$ and
thus correspond to stable states, while the root $v=0$ maximises $G$
and thus corresponds to an unstable state. Stable states represent
the coexisting phase of liquid and vapour, with pressure given by
\begin{eqnarray}
P_2 = P_c(1+4t) = P_c \left( 1-4\frac{T_c-T}{T_c}\right).
\end{eqnarray}
The liquid phase is more stable when $P_2<P<P_c$, and the vapour
phase more stable when $P<P_2$.

\subsection{The thermodynamic limit}
\label{sec:3.4}

The existence of the instability in the van der Waals system studied
above is related to the fact that in the canonical distribution the
volume $V$ of the system is held fixed, whereas in a real gas
\textit{volume fluctuations are significant in the vicinity of the
critical point}. In other words, the canonical distribution does not
provide a completely accurate physical description of the
vapour-liquid equilibrium. Therefore, as in the case of an ideal
gas, we consider the pressure-temperature distribution, with the
corresponding partition function
\begin{eqnarray}
Z(\alpha,\beta) = \frac{1}{b}\int_{bN}^\infty z(\beta,V)
\exp(-\alpha V)\rd V \label{eq:3.84}
\end{eqnarray}
wherein the volume fluctuation is integrated out. Recall that
$b=\frac{2}{3}\pi d^3$ represents the smallest volume each molecule
can occupy. Hence, the random variable $V$ representing the total
volume ranges from $bN$ to infinity. When the canonical partition
function (\ref{eq:3.76}) is substituted into (\ref{eq:3.84}), the
resulting integral does not admit an elementary analytical
expression. Nevertheless, in the thermodynamic limit $N\to\infty$ we
can implicitly determine the potential $\psi(\alpha,\beta) =
N^{-1}\ln Z(\alpha,\beta)$ by the method of steepest descent.

We proceed as follows. First we write the integrand in
(\ref{eq:3.84}) as
\begin{eqnarray}
\exp(-\alpha V)z(\beta,V) = \exp[Ng(\alpha,\beta,\hat{v})] ,
\end{eqnarray}
where $\hat{v}\equiv V/N$ and
\begin{eqnarray}
g(\alpha,\beta,\hat{v})=1-\alpha\hat{v}+\ln\left[\Big(\frac{2\pi
m} {\beta h^{2}}\Big)^{\frac{3}{2}}(\hat{v}-b)\right]+\frac{\beta
a} {\hat{v}}. \label{eq:3.86}
\end{eqnarray}
In deriving (\ref{eq:3.86}) we have used the Stirling formula
$\ln N!\simeq N\ln N-N$. It should be evident from (\ref{eq:3.80})
that
\begin{eqnarray}
G=-\beta^{-1} g(\alpha,\beta,\hat{v})
\end{eqnarray}
is the Gibbs free energy. Also, note that $g(\alpha,\beta,\hat{v})$
must have at least one maximum in the range $\hat{v}\in[b,\infty)$
corresponding to the minimum Gibbs free energy, because
$g(\alpha,\beta,\hat{v}) \to-\infty$ in the limits $\hat{v}\to b$
and $\hat{v}\to\infty$. The value of $\hat{v}$ at which
$g(\alpha,\beta,\hat{v})$ is maximised therefore determines the
equation of state (\ref{eq:3.77}) for the canonical distribution.
However, in the $P$-$T$ distribution the volume is a random
variable, hence we must take its expectation to obtain the equation
of state:
\begin{eqnarray}
\langle\hat{v}\rangle = -\frac{1}{N}\frac{\partial\ln Z(\alpha,
\beta)}{\partial \alpha}, \label{eq:3.87}
\end{eqnarray}
where $\langle\hat{v}\rangle$ denotes the expected volume per
particle in the $P$-$T$ distribution characterised by the density
function $Z^{-1}(\alpha,\beta) \exp[Ng(\alpha,\beta, \hat{v})]$.

Applying the change of variable $V\to\hat{v}$, the partition
function (\ref{eq:3.84}) can be written in the form
\begin{eqnarray}
Z(\alpha,\beta) = \frac{N}{b}\int_b^\infty \exp\left[ N g(\alpha,
\beta,\hat{v})\right] \rd \hat{v}.
\end{eqnarray}
Recall that we are interested in the thermodynamic potential per
particle in the thermodynamic limit:
\begin{eqnarray}
\psi(\alpha, \beta) = \lim_{N\to\infty} N^{-1}\ln Z(\alpha,\beta).
\end{eqnarray}
Using the method of steepest descent, we find that
$\psi(\alpha,\beta)$ in this limit is given by
\begin{eqnarray}
\psi(\alpha,\beta) = g(\alpha,\beta,\bar{v}) = -\alpha \bar{v} +
\ln z(\beta,\bar{v}), \label{eq:3.89}
\end{eqnarray}
where $\bar{v}=\bar{v}(\alpha,\beta)$ is the function of $\alpha$
and $\beta$ which maximises $g(\alpha,\beta,\hat{v})$. Since
$\bar{v}$ minimises the Gibbs free energy, it is the solution of the
van der Waals equation of state. Although the exact functional form
of $\bar{v}(\alpha,\beta)$ is not at our disposal owing to the cubic
nature of the equation of state, we can nonetheless determine the
exact expression for the scalar curvature in terms of the variables
$\beta$ and $\bar{v}$. Before we proceed, however, we first
establish the following result.

\begin{prop}\label{prop:6}
The thermal expectation value of the volume $\hat{v}$ per particle
in the $P$-$T$ distribution is given by $\bar{v}$, that is,
$\langle\hat{v}\rangle=\bar{v}$.
\end{prop}

{\it Proof}. Differentiating (\ref{eq:3.89}) and using the
chain rule we find
\begin{eqnarray}
\frac{\partial\psi}{\partial\alpha}=-\bar{v}+\Big(-\alpha+
\frac{\partial\ln z}{\partial\bar{v}}\Big) \frac{\partial\bar{v}}
{\partial\alpha}.
\end{eqnarray}
However, by definition $\bar{v}$ maximises $g(\alpha,\beta,
\hat{v})$ so that
\begin{eqnarray}
\frac{\partial g}{\partial\bar{v}}=-\alpha+ \frac{\partial\ln z}
{\partial\bar{v}}=0, \label{eq:3.145}
\end{eqnarray}
and hence
\begin{eqnarray}
\frac{\partial\psi}{\partial\alpha}=-\bar{v}. \label{eq:3.146}
\end{eqnarray}
On the other hand, from (\ref{eq:3.87}) we have
$\langle\hat{v}\rangle = -\partial\psi/\partial\alpha$, and thus
$\langle\hat{v}\rangle=\bar{v}$. \hspace*{\fill}
$\square$

\subsection{Geometry of the van der Waals manifold}
\label{sec:3.5}

As we have just indicated, the functional form of $\bar{v}(\alpha,
\beta)$ is unknown. Nevertheless, we can implicitly determine
expressions for $\partial\bar{v}/\partial\alpha$, $\partial\bar{v}/
\partial\beta$, and so on, in the following manner. First, define
$\Omega$ by
\begin{eqnarray}
\Omega \equiv - \frac{\partial g}{\partial\bar{v}} = \alpha -
\frac{1}{\bar{v}-b} + \beta\frac{a}{\bar{v}^{2}} . \label{eq:3.147}
\end{eqnarray}
Since $\hat{v}=\bar{v}$ maximises $g(\alpha,\beta,\hat{v})$ we have
by definition the relation $\Omega=0$. This, however, is just the
equation of state for the van der Waals gas. Now, consider the total
derivative of $\Omega$:
\begin{eqnarray}
{\rm d}\Omega = \frac{\partial \Omega} {\partial\alpha}{\rm d}\alpha
+ \frac{\partial\Omega}{\partial\beta} {\rm d}\beta +
\frac{\partial\Omega} {\partial\bar{v}}{\rm d}\bar{v}.
\label{eq:3.148}
\end{eqnarray}
Since $\rd\Omega=0$ it follows that
\begin{eqnarray}
{\rm d}\bar{v} &=& -\left[ \frac{\partial\Omega}{\partial\alpha}
\Big/ \frac{\partial\Omega}{\partial\bar{v}}\right] {\rd} \alpha -
\left[\frac{\partial\Omega} {\partial\beta}\Big/
\frac{\partial\Omega} {\partial\bar{v}}\right] {\rd}\beta
\nonumber
\\ &=& \Big(\frac{\partial\bar{v}}{\partial\alpha}\Big)_\beta
{\rd}\alpha + \Big(\frac{\partial\bar{v}}{\partial\beta}
\Big)_\alpha {\rd}\beta , \label{eq:3.149}
\end{eqnarray}
where we have used the general identity:
\begin{eqnarray}
\Big(\frac{\partial\alpha}{\partial\beta}\Big)_\gamma
\Big(\frac{\partial\beta}{\partial\gamma}\Big)_\alpha
\Big(\frac{\partial\gamma}{\partial\alpha}\Big)_\beta = -1.
\label{eq:3.150}
\end{eqnarray}
On the other hand, from (\ref{eq:3.147}) we have the relations
\begin{eqnarray}
\frac{\partial\Omega}{\partial\alpha}=1,\quad
\frac{\partial\Omega}{\partial\beta}=\frac{a}{\bar{v}^{2}}, \quad
{\rm and}\quad \frac{\partial\Omega}{\partial\bar{v}}=
\frac{1}{(\bar{v}-b)^{2}}-\frac{2a\beta}{{\bar v}^{3}} .
\label{eq:3.151}
\end{eqnarray}
Therefore, substituting these into (\ref{eq:3.149}) we deduce that
\begin{eqnarray}
\frac{\partial\bar{v}}{\partial\alpha} = \frac{1}{D}\quad {\rm and}
\quad \frac{\partial\bar{v}}{\partial\beta} = \frac{1}{D}
\frac{a}{\bar{v}^{2}}, \label{eq:3.152}
\end{eqnarray}
where $D$ is defined by
\begin{eqnarray}
D(\alpha,\beta) = \frac{2a\beta}{{\bar v}^{3}} -
\frac{1}{(\bar{v}-b)^{2}} . \label{eq:3.153}
\end{eqnarray}

The derivatives of $\bar{v}$ with respect to the parameters $\alpha$
and $\beta$ are required in order to determine the components of the
Fisher-Rao metric on the van der Waals manifold. Specifically we
obtain the following result:

\begin{prop}\label{prop:7}
In terms of the
pressure-temperature coordinates $(\alpha,\beta)$ the Fisher-Rao
metric on the van der Waals manifold is given by
\begin{eqnarray}
{G}_{ij} = \frac{1}{D} \left( \begin{array}{cc} -1 & -a/\bar{v}^{2}
\\ -a/\bar{v}^{2} & \ \ \frac{3}{2}\beta^{-2} - (a/\bar{v}^{2})^{2}
\\ \end{array}  \right) . \label{eq:3.154}
\end{eqnarray}
In particular, in the ideal gas limit $a\to0$ and $b\to0$,
the metric {\rm (\ref{eq:3.154})} reduces to the metric {\rm
(\ref{eq:3.52})} for the ideal gas.
\end{prop}

{\it Proof}. The components of the metric are determined by the
matrix $\partial_i\partial_j\psi(\alpha,\beta)$. We have, in
Proposition~\ref{prop:6}, established that $\partial\psi/
\partial\alpha=-\bar{v}$, and, using (\ref{eq:3.145}),we have
\begin{eqnarray}
\frac{\partial\psi}{\partial\beta}=-\alpha\frac{\partial\bar{v}}
{\partial\beta}+\frac{\partial\ln z}{\partial\bar{v}}
\frac{\partial\bar{v}}{\partial\beta} + \frac{\partial\ln
z}{\partial\beta}=\frac{\partial\ln z}{\partial\beta},
\label{eq:3.155}
\end{eqnarray}
Therefore, we obtain
\begin{eqnarray}
\fl \frac{\partial^2\psi}{\partial\alpha^2} =
-\frac{\partial\bar{v}} {\partial\alpha}, \qquad
\frac{\partial^2\psi}{\partial\alpha
\partial\beta} = - \frac{\partial\bar{v}}{\partial\beta}, \qquad
{\rm and} \qquad \frac{\partial^2\psi}{\partial\beta^2}=
\frac{\partial^2\ln z}{\partial\beta^2}+ \frac{\partial^2\ln
z}{\partial \beta\partial\bar{v}}
\frac{\partial\bar{v}}{\partial\beta}, \label{eq:3.156}
\end{eqnarray}
whence the desired expression for the metric follows from the
formula (\ref{eq:3.76}) for the canonical partition function. In the
ideal gas limit $a\to0$ and $b\to0$, we have $D\to\bar{v}$. However,
from the ideal gas equation of state we have $\bar{v}=\alpha^{-1}$,
hence we recover (\ref{eq:3.52}) at once in this limit.
\hspace*{\fill} $\square$

\begin{figure}[t]
\begin{center}
  \includegraphics[scale=1.0]{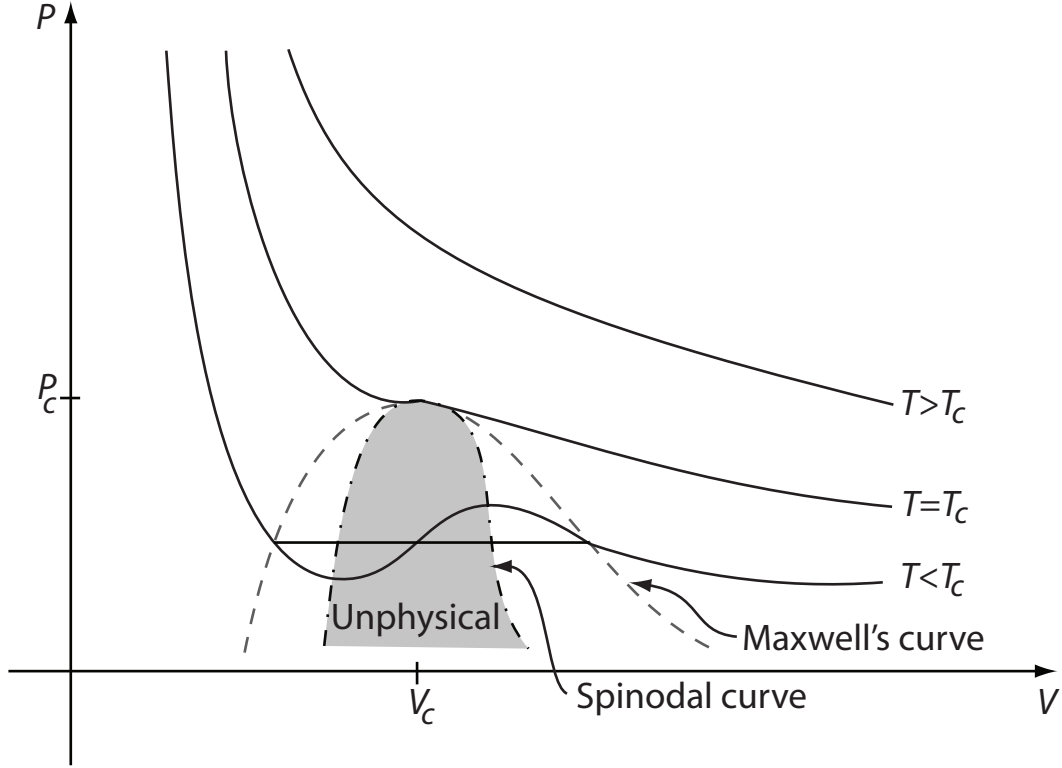}
  \caption{\label{fig:3.5}  Schematic illustration of equations
 of state for
 van der Waals gas. The scalar curvature on the parameter
 space diverges along the spinodal boundary which envelopes the
 unphysical region. The critical point is that where the spinodal curve is
 tangent to the Maxwell equal area boundary. The divergence of the
 curvature along the spinodal boundary may be interpreted as
 `preventing', in some sense, entry into
 the unphysical domain in the phase diagram.
 }
\end{center}
\end{figure}

To describe the geometry of the van der Waals gas it will be
convenient to introduce the concept of a \textit{spinodal curve}. In
general a spinodal curve consists of the points in the thermodynamic
phase space at which the second derivative of the free energy with
respect to an order parameter vanishes. For a gas of interacting
molecules, the mean volume $\bar{v}$ constitutes the order parameter
of the system, and the vanishing of the second derivative of $\ln
z(\beta,\bar{v})$ with respect to $\bar{v}$ thus determines the
spinodal curve in the phase diagram. For the van der Waals gas, by
(\ref{eq:3.145}), we have the relation
\begin{eqnarray}
\frac{\partial\alpha}{\partial\bar{v}} = -\frac{\partial^2\ln z}
{\partial\bar{v}^2} = 0 \label{eq:3.157}
\end{eqnarray}
that determines the spinodal curve. In other words, the locus of
points at which the derivative of pressure
with respect to volume vanishes for some temperature determines the
spinodal curve.
This is schematically illustrated for the van der Waals equation
in Figure~\ref{fig:3.5}.

It is evident from Figure~\ref{fig:3.5} that the region in the phase
diagram enveloped by the spinodal curve is unstable because
$(\partial P/\partial V)_T>0$ in this region, i.e. the pressure
increases with increasing volume. The spinodal curve thus forms the
boundary of a semi-stable region in the phase diagram. In view of
the first relation in (\ref{eq:3.152}), the spinodal curve is
determined by the condition $D=0$. On the other hand, from the
expression (\ref{eq:3.154}) for the Fisher-Rao metric on the van der
Waals manifold we see that each component of the metric ${G}_{ij}$
as well as its determinant $-3/(2\beta^2D)$ is singular along the
spinodal curve. Is this singular behaviour merely due to the
specific choice of coordinates or is it an intrinsic feature of the
van der Waals manifold? We can answer this question by calculating
the scalar curvature of the manifold. The exact expression for the
curvature is as follows.

\begin{prop}\label{prop:8} The scalar curvature of the van
der Waals manifold is given by
\begin{eqnarray}
R = \frac{4}{3D^2}\left(\frac{a\beta}{\bar{v}}\right)\left(
\frac{a\beta}{\bar{v}^{3}} - D \right). \label{eq:3.158}
\end{eqnarray}
In particular, $R$ diverges along the entire spinodal curve
specified by $D=0$, which includes the critical point
$(P_c,V_c,T_c)$. The scalar curvature vanishes in the ideal gas
limit for which $a\to0$ and $b\to0$.
\end{prop}

{\it Proof}. Since we have chosen the canonical parametrisation
$(\alpha, \beta)$, we can use the determinant in (\ref{eq:3.16}) to
calculate the curvature. To compute the entries in the determinant
we differentiate (\ref{eq:3.156}) and use the chain rule, together
with (\ref{eq:3.152}), thereby obtaining
\begin{eqnarray}
\psi_{111}&=&\frac{2}{D^3}\Big(\frac{1}{(\bar{v} -b)^3} -3
\frac{a\beta}{\bar{v}^4}\Big)\equiv X, \quad \psi_{112}= \frac{a}
{\bar{v}^2}X+\frac{2a}{\bar{v}D^2} \nonumber \\ \psi_{122}&=&
\frac{a^2}{\bar{v}^4}\Big(X+\frac{4}{\bar{v}D^2}\Big), \quad
\psi_{222}= \frac{a^3}{\bar{v}^6}\Big(X+\frac{6}{\bar{v}D^2}
\Big)-\frac{3}{\beta^3}. \label{eq:3.159}
\end{eqnarray}
Substituting these results into (\ref{eq:3.16}) we obtain, after
some algebra, the desired expression in (\ref{eq:3.158}). The fact
that the curvature diverges along the spinodal curve $D=0$ is
evident from the expression (\ref{eq:3.158}). Also, from the
definition (\ref{eq:3.157}) of the spinodal curve and the condition
(\ref{eq:3.63}) for the critical point, it is clear that the
critical point lies on the spinodal curve. The vanishing of the
curvature in the ideal gas limit $a,b\to0$ is also evident from the
expression in (\ref{eq:3.158}). \hspace*{\fill} $\square$

The van der Waals manifold possesses the structure of a Riemann
surface over a planar base space with coordinates $(\alpha,\beta)$,
branched around the singularities specified by the spinodal curve.
Now, suppose we slowly change the variables $(\alpha,\beta)$ along a
closed contour $C$ in the planar base space. Then, the lifted curve
in the van der Waals manifold does not, in general, return to the
same sheet (i.e. to the same thermodynamic state) if $C$ encloses
the critical point or crosses the spinodal curve, while Maxwell's
relation ensures that an infinitesimal closed contour enclosing no
singularities is thermodynamically trivial. Thus, the presence of
singularities may give rise to changes in the thermodynamic state
$\bar{v}$ of the system upon following a closed contour in the
parameter base space which encloses a point of divergency.
Conversely, if a closed contour in the parameter space does not
enclose the critical point or cross the spinodal curve, then the
corresponding curve in ${\mathfrak M}$ is closed and thus gives rise
to a well-defined holonomy. This leads naturally to the following
open problem: What is the physical interpretation or relevance of
the holonomy (analogue of the geometric phase in quantum mechanics)
in classical statistical mechanics?

\begin{figure}[t]
\begin{center}
  \includegraphics[scale=0.8]{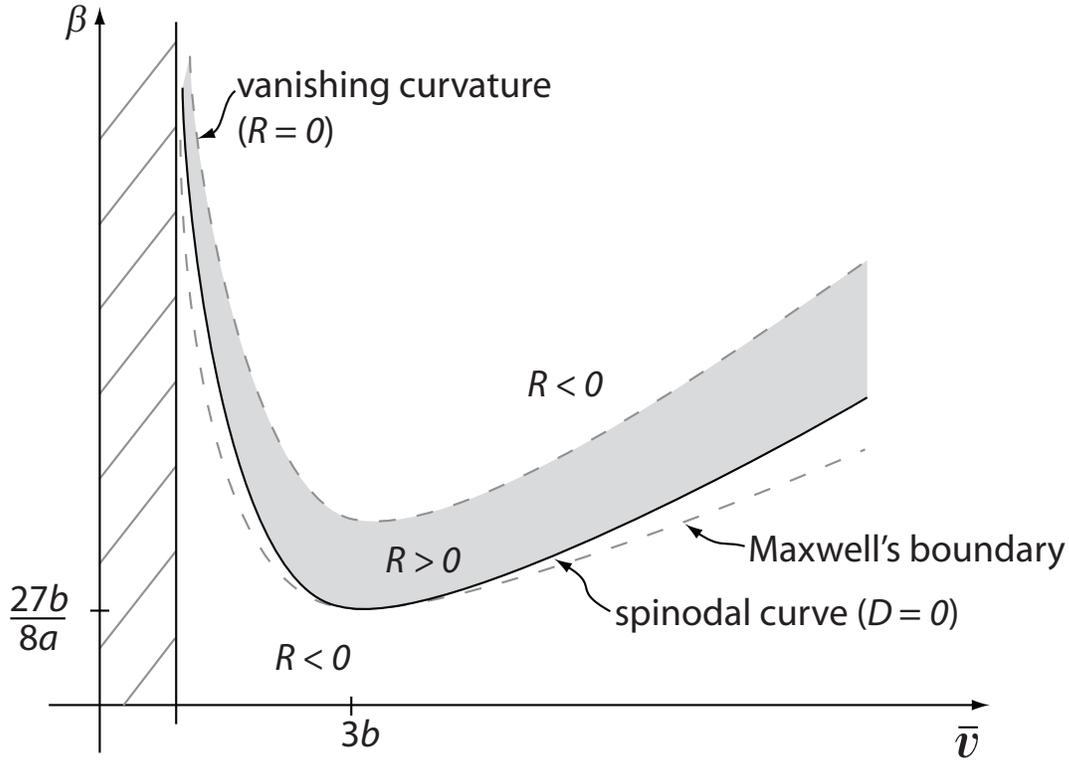}
  \caption{\label{fig:3.6} Geometric phase diagram for the van der
  Waals gas. The gas is divided into positive and negative curvature
  phases by the vanishing curvature $R=0$ curve. The change of phase
  along $R=0$ is analytic, while the curvature exhibits singular
  behaviour along the spinodal curve $D=0$.
 }
\end{center}
\end{figure}

Finally we note that the scalar curvature $R$ on the van der Waals
manifold vanishes along the curve specified by
\begin{eqnarray}
\beta = \frac{\bar{v}^{3}}{a(\bar{v}-b)^2} , \label{eq:3.160}
\end{eqnarray}
and its sign changes smoothly from positive to negative
as one decreases the temperature in the $(\bar{v},\beta)$
plane. The sign of the scalar curvature in the $(\bar{v},\beta)$
plane and its relation to the spinodal and Maxwell boundaries are
schematically illustrated in Figure~\ref{fig:3.6}. One might refer
to the smooth change in the sign of the scalar curvature in the
phase diagram as a \emph{geometric phase transition}. Unlike the
conventional phase transitions associated with singular
behaviour, however, such geometric phase transitions are not
associated with any divergence. There are indications that the
scalar curvature can be viewed as a measure of stability of the
system. However, the precise correspondence between physical
characteristics of the system and properties of the curvature,
apart from the singular behaviour along the spinodal boundary,
remains an open research problem.

\section{Bibliographical notes}
\label{sec:4}

When the first author was invited to write a survey article on a
topic of interest, it seemed appropriate to utilise this opportunity
by briefly reviewing some applications of information geometry in
physics. This reflects the fact that interest in this area has
continued to grow. Two international conferences on applications of
information geometry have been held in recent years, while diverse
new applications continue to emerge---for example, in the area of
shape recognition in computer science (Maybank 2005, Peter \&
Rangarajan 2006), in connection with out of equilibrium measures
(Crooks 2007), in the characterisation of quantum phase transitions
(Zanardi \textit{et al}. 2007), in various mathematical extensions
(e.g., Cena \& Pistone 2007), or in black hole thermodynamics
(Ruppeiner 2008). The application of information geometry to
statistical physics, however, has generated a vast amount of
literature, and it seemed neither feasible nor helpful to cover all
the results which have emerged in this area. What seemed more
appropriate was to focus attention upon one specific topic that
nevertheless incorporates all the essential ingredients. It also
appeared desirable to explain the basic ideas of information
geometry and its emergence from probabilistic and statistical
considerations in a language accessible to a graduate student in
theoretical physics. For these reasons we begin the paper with a
some rather elementary background material, and then consider its
application to the theory of vapour-liquid equilibrium. This leads
to the analysis of the van der Waals gas model, which, in our
opinion, not only embodies all the essential features of a phase
transition in statistical mechanics but also admits an elegant
geometric characterisation. To keep the exposition at a fairly
elementary level, we have excluded \textit{ad hoc} citations from
the main text so as to avoid impeding the continuity of the
exposition. Instead, references to the literature are presented more
informally in these bibliographical notes.

To the authors' best knowledge, the use of geometric methods in
statistical analysis was first introduced by P.~C.~Mahalanobis, the
founder of the Indian Statistical Institute and also the founding
editor of Sankhy\={a} (The Indian Journal of Statistics), in the
early 1930's (Mahalanobis 1930, 1936). Mahalanobis was a physicist
and statistician who taught relativity and wrote an introduction to
the translation by M.~Saha of Minkowski's work on relativity. His
articles on relativity, written jointly with S.~N.~Bose were
published by the Calcutta University. He introduced a measure of
mutual separation in the study of statistical data arising from
anthropometric measurements. An alternative measure of separation
was subsequently introduced by Bhattacharyya (1943, 1946), and was
defined in Section~\ref{sec:1} as the Bhattacharyya spherical
distance between two probability densities.

The geometric description of the parameter-space manifold was
initiated at around the same time by Rao (1945, 1947, 1954). The
seminal paper by Rao (1945) is significant in two respects: on the
one hand, it introduced the so-called Cram\'er-Rao inequality as a
lower bound for the variance, while on the other hand it pointed out
that the information measure introduced previously by Fisher (1925)
defines a Riemannian metric on the parameter-space manifold of a
statistical model. Rao then proposed the associated geodesic
distance as a measure of dissimilarity between probability
distributions. The formulation considered by Rao was based on the
Hilbert space embedding $p_\theta(x) \to l_\theta(x) =\surd
p_\theta(x)$. As a consequence, many of the constructions in Rao
(1945) bear a formal resemblance to the geometric formulation of
quantum mechanics, developed by physicists some time later in the
1980's and 1990's. Concurrently with Rao's work on the application
of geometry in statistics, Jeffreys (1946) also introduced the
concept of uninformative priors based on the use of a Riemannian
metric.

It is worth noting incidentally that the inner product of
probability measures via the square-root embedding was introduced
earlier by Hellinger (1909) in connection with unitary invariants of
self-adjoint operators in Hilbert space. The distance measure of
Hellinger was subsequently extended by von Neumann and Kakutani (see
Kakutani 1946), who introduced an inner product of probability
measures in an abstract measure-theoretic context and applied this
to investigate equivalence and orthogonality relations between
product measures. The Kakutani inner product was used by Brody
(1971) to provide a simple proof of the Gaussian dichotomy theorem,
and has also been extended by Bures (1969) in the context of
operator algebras.

Interest in the application of geometrical techniques to statistical
inference appear to have somewhat diminished subsequently, but
reemerged with the appearance of Efron's seminal paper (1975) on
information loss and statistical curvature. Efron considered the
logarithmic embedding $p_\theta(x) \to l_\theta(x) =\ln p_\theta(x)$
and demonstrated that in the space of log-likelihood density
functions the curvature of the curve $l_\theta$ measures the
deviation of the density function from the exponential family.
Furthermore, the squared statistical curvature was shown to
determine the loss of information resulting from the use of the
maximum likelihood estimator for the unknown parameter $\theta$.

The work of Efron (1975, 1978)---and to some extent that of
\v{C}encov (1982) who showed that the Fisher information metric is
unique under certain assumptions including invariance---evoked
considerable interest in the geometrical approach to asymptotic
inference and related topics in statistics. Numerous research papers
(for example, Atkinson \& Mitchell 1981), as well as review papers
(for example, Kass 1989) and monographs (for example, Amari 1985,
Amari \textit{et al}. 1987, Murray \& Rice 1993, and Amari \&
Nagaoka 2000), were subsequently published.

In parallel with these developments in statistics, the application
of information geometry to the description of the equilibrium
properties of thermodynamic systems was considered by a number of
authors. One of the initiators was Ingarden (1981), who considered
various Banach space embeddings of probability density functions and
their relation to the concepts of entropy and Fisher information,
and suggested their application to statistical physics. The works of
Ingarden and his collaborators led to the establishment of a Polish
School of researchers systematically investigating various geometric
aspects of physical systems described by equilibrium distributions.
Janyszek \& Mruga{\l}a (1989a), for instance, clarified the relation
of the Fisher-Rao geometry to contact geometry (the latter
characterises the geometry of Legendre transformations), and
investigated the physical interpretation of the metric tensor for a
system of gas molecules characterised by the pressure-temperature
distribution. Janyszek \& Mruga{\l}a (1989b) calculated the scalar
curvatures of the parameter space manifolds of the one-dimensional
Ising model in the thermodynamic limit and of the mean-field model.
Janyszek \& Mruga{\l}a (1990) also extended information geometric
analysis to the investigation of the stability of ideal quantum
gases. Some of these ideas were further extended by others in the
context of various spin models in statistical mechanics (see, for
example, Janke \textit{et al}. 2002, Brody \& Ritz 2003, Johnston
\textit{et al}. 2003).

An independent line of investigation on the various geometric
properties of thermodynamic state spaces was initiated by Weinhold
(1975) and also by Ruppeiner (1979). Weinhold proposed the existence
of a metric structure for the thermodynamic state space arising from
empirical laws of  thermodynamics. This line of thinking, which led
to the notion of the so-called \textit{thermodynamic length}, was
extended in a variety of ways by various authors (see, for example,
Salamon \& Berry 1983, Schl\"ogl 1985, Mruga{\l}a \textit{et al}.
1990).

Ruppeiner went a step further and considered the second derivative
of the entropy, discussed briefly here in Section~\ref{sec:1.5}, as
a Riemannian metric on the thermodynamic state space. The metric
considered by Ruppeiner, based upon the Shannon-Wiener entropy,
agrees with Rao's entropy derivative metric (cf. Rao 1984), and is
also related to the Fisher-Rao metric through a Legendre
transformation. The key idea is that the convexity of entropy
implies the positive-definiteness of the associated Hessian matrix,
which can therefore be used to define a Riemannian metric on the
thermodynamic state space. The simplest system to consider in this
respect is naturally that of a noninteracting gas of classical
particles. This was investigated by Ruppeiner (1979), and also by
Mijatovi\'{c} \textit{et al}. (1987). Geometric aspects of various
other physical systems have also been investigated along these lines
(Ruppeiner 1990, 1991). For a comprehensive bibliography on this
topic, see the reference list in the review article by Ruppeiner
(1995).

Ingarden \& Tamassy (1993) also applied an entropic measure of
divergence to explain the thermodynamic arrow of time. While the
infinitesimal form of the entropic measure of divergence (relative
entropy) gives rise to a Riemannian structure characterised in
general by the Burbea-Rao metric (\ref{eq:1.85}), the metrics
arising from entropies generally possess Finslerian structure. In
particular, the Finsler metric arising from relative entropy is not
symmetric. The idea of Ingarden \& Tamassy (1993) consists in
exploiting the lack of symmetry in such metrics to explain the
thermodynamic arrow of time, without the introduction of
dissipation.

An important application of information geometry to the properties
of renormalisation group flow in statistical physics---motivated in
part by the observation of Dawid (1975) that Efron's results might
be represented more concisely in terms of Hilbert space
geometry---was proposed by Brody (1987). Closely related ideas were
developed further by O'Connor \& Stephens (1993), Dolan (1998),
Brody \& Ritz (1998), and Brody (2000). (See also Di\'osi \textit{et
al}. (1984) for an alternative approach to analysis of
renormalisation group flow using Rao's entropy derivative metric.)

The ideas of information geometry have also been extended to the
quantum domain by replacing the density functions of classical
probability theory by density matrices. Substantial work has been
done on quantum information geometry (for example, Petz \& Sudar
1996, Uhlmann 1996, Grasselli \& Streater 2001, Petz 2002, Streater
2004, Jen\v{c}ov\'{a} \& Petz 2006, Gibilisco \& Isola 2007, and
Gibilisco \textit{et al}. 2007), as well as its application to
quantum statistical inference (Brody \& Hughston 1998,
Barndorff-Nielsen \& Gill 2000).

Turning more specifically to the subject matter of the present
review, as indicated above, the spherical distance (\ref{eq:1.7})
representing the dissimilarity of probability densities was
introduced by Bhattacharyya (1943), and the concept of a statistical
manifold represented by the metric (\ref{eq:1.23}) was introduced by
Rao (1945). The uniqueness of the Einstein metric on a complex
projective space was conjectured by Calabi, and later proven by Yau
(1977). The implication of this result in quantum mechanics---that
the solution to the vacuum Einstein equation in the space of pure
quantum states determines transition probabilities---was pointed out
to the first author by G.~W.~Gibbons in the late 1990's. The
relevance of the projective space and the associated metric
(\ref{eq:1.47}) to statistical mechanics was demonstrated by Brody
\& Hughston (1999).

The expression in Proposition~\ref{prop:2} for the scalar curvature
in terms of the determinant of a $3\times3$ matrix, valid for the
two-dimensional statistical manifold associated with a canonical
density function, is given in Janyszek \& Mruga{\l}a (1989b). The
fact that the statistical manifold associated with the normal
density function possesses constant negative curvature, shown in
equation (\ref{eq:1.57}), was observed by Amari (1982). However, the
significance of the scalar curvature (or in fact the Riemann tensor
itself) in statistical analysis remains somewhat obscure.

A survey article by Burbea (1986) deals systematically with
expressions for geodesic curves associated with a number of standard
density functions used in statistics (see also the article by Rao in
Amari \textit{et al}. 1987). This includes, in particular, the
distance between Gaussian density functions as given in
(\ref{eq:1.69}) and (\ref{eq:1.71}). Analogous results for
gamma-distributed densities have been calculated in some detail by
Burbea \textit{et al}. (2002).

The fact that the leading order term in the Taylor expansion of the
relative entropy of a neighbouring pair of parametric density
functions gives rise to the Fisher-Rao metric was observed by
Ingarden (1981). A more detailed and thorough analysis was given by
Burbea \& Rao (1984), and constitutes the basis for the discussion
in Section~\ref{sec:1.5}. The specific form of entropy defined in
(\ref{eq:1.80}), sometimes referred to as the $\alpha$-order
entropy, was introduced by Havrad \& Charv\'{a}t (1967) in the
context of quantifying classification schemes. See also Burbea \&
Rao (1982a, 1982b) for details concerning various properties of this
entropy. The use of the $\alpha$-order entropy in statistical
mechanics has been proposed by Tsallis (1988).

In Section~\ref{sec:2} we considered the information geometry of the
pressure-temperature distribution representing the equilibrium state
of a gas of noninteracting particles (classical ideal gas). The
flatness of the information manifold of a classical ideal gas was
pointed out by Ruppeiner (1979) using the entropy derivative metric.
The solution to the associated geodesic equations, in the form
expressed in Proposition~\ref{prop:5}, does not seem to appear
elsewhere. (An alternative representation of the geodesics appears
in Mijatovi\'{c} \textit{et al}. 1987.)

In Sections~\ref{sec:3.1}$\sim$\ref{sec:3.4} we have provided a
brief account of the classical theory of the van der Waals gas, as a
background for the subsequent geometric description. Our exposition
follows closely the classic treatise of Mayer \& Mayer (1940). There
is a series of inspiring papers by Kac \textit{et al}. (1963),
Uhlenbeck \textit{et al}. (1963), Hemmer \textit{et al}. (1964), and
also Hemmer (1964), analysing the vapour-liquid equilibrium of the
van der Waals gas in great detail. These papers extend the earlier
work of Kac (1959), which provides a method for determining the
partition function of interacting gas molecules.

Other related work on systems of interacting gas molecules includes
the following: Tonks (1936) determined the equations of state for
gases composed of hard elastic spheres with finite radius. Van Hove
(1950) calculated the free energy of a system of molecules with
nonvanishing incompressible radii, interacting according to a finite
range force. He showed that in one dimension the system exhibits no
phase transition. Lebowitz \& Percus (1963) studied the properties
of the correlation functions. Van Kampen (1964) showed that a gas of
molecules with hard sphere repulsive forces and long-range
attractive interactions exhibits condensation, and calculated the
density fluctuations. Rigourous bounds for the free energy of the
van der Waals gas were derived by Lebowitz \& Penrose (1966).

Detailed analyses of the curvature of some of these classical
systems of interacting molecules were presented by Ruppeiner \&
Chance (1990). The geometry of the van der Waals gas associated with
the entropy derivative metric is considered in papers by Di\'osi \&
Luk\'{a}cs (1986) and in Di\'osi {\it et al}. (1989), wherein the
authors determine the scalar curvature using the density and
temperature as coordinates. Using these coordinates, they have also
shown that on this statistical manifold there exists no solution to
the Killing equations (i.e. no vector field such that the associated
flow preserves geodesic distances).

The description of the geometry of the van der Waals manifold
presented in Section~\ref{sec:3.5} follows closely the analysis
outlined by Janyszek (1990) and also by Brody \& Rivier (1995). In
particular, the expressions for the metric tensor in
Proposition~\ref{prop:7} and for the scalar curvature in
Proposition~\ref{prop:8} were derived by Brody \& Rivier (1995), who
also suggested that the curvature of the statistical manifold might
play a role in statistical mechanics analogous to that of geometric
phases in quantum mechanics. This remains an open issue, although
recent work on quantum phase transitions indicate that there is
indeed a close analogy between these two concepts.

Finally, the present authors regret that owing to the huge volume of
literature on this subject there are many other valuable
contributions which have not been mentioned in these brief
biographical notes.



\vspace{0.5cm}

\begin{thebibliography}{999}

\bibitem{amari1} {\sc Amari,~S.} (1982) Differential
geometry of curved exponential families---curvatures and information
loss. {\em The Annals of  Statistics} \textbf{10}, 357-385.

\bibitem{amari2}
{\sc Amari,~S.} (1985) {\em Differential-geometrical methods in
statistics}. Lecture Notes in Statistics \textbf{28} (New York:
Springer-Verlag).

\bibitem{amari3}
{\sc Amari,~S., Barndorff-Nielsen,~O.~E., Kass,~R.~E.,
Lauritzen,~S.~L. \& Rao,~C.~R.} (1987) {\em Differential geometry in
statistical inference}. Institute of Mathematical Statistics Lecture
Notes. Monograph Series \textbf{10}. (Hayward, CA: Institute of
Mathematical Statistics).

\bibitem{amari4} {\sc Amari,~S. \& Nagaoka,~H.} (2000) {\em Methods
of Information Geometry}. {\em AMS Translations of Mathematical
Monograph} \textbf{191} (Oxford: Oxford University Press).

\bibitem{atkinson}
{\sc Atkinson,~C. \& Mitchell,~A.~F.~S.} (1981) Rao's distance
measure. {\em Sankhy\={a}} \textbf{43}, 345-365.

\bibitem{barndorff}
{\sc Barndorff-Nielsen,~O.~E. \& Gill,~R.~D.} (2000) Fisher
information in quantum statistics. {\em Journal of Physics}
A\textbf{33}, 4481-4490.

\bibitem{Bhat2} {\sc Bhattacharyya,~A.} (1943)
On a measure of divergence between two statistical populations
defined by their probability distributions. {\em Bulletin of the
Calcutta Mathetical Society} \textbf{35}, 99-109.

\bibitem{Bhat3} {\sc Bhattacharyya,~A.} (1946)
On a measure of divergence between two multinomial populations. {\em
Sankhy\={a}} \textbf{7}, 401-406.

\bibitem{brody0}
{\sc Brody,~D.~C.} (2000) Differential renormalisation
flow in random lattice gauge theories. {\em Physics Letters}
B\textbf{485}, 422-428.

\bibitem{brody1}
{\sc Brody,~D.~C. \& Hughston,~L.~P.} (1998) Statistical geometry in
quantum mechanics. {\em Proceedings of the Royal Society London}
A\textbf{454}, 2445-2475.

\bibitem{brody2}
{\sc Brody,~D.~C. \& Hughston,~L.~P.} (1999) Geometrisation of
statistical mechanics. {\em Proceedings of the Royal Society London}
A\textbf{455}, 1683-1715.

\bibitem{brody3}
{\sc Brody,~D.~C. \& Ritz,~A.} (1998) On the symmetry of real-space
renormalisation. {\em Nuclear Physics} B\textbf{522}, 588-604.

\bibitem{brody4}
{\sc Brody,~D.~C. \& Ritz,~A.} (2003) Information geometry of finite
Ising models. {\em Journal of Geometry and Physics} \textbf{47},
207-220.

\bibitem{brody5}
{\sc Brody,~D.~C. \& Rivier,~N.} (1995) Geometrical aspects of
statistical mechanics, {\em Physical Review} E\textbf{51},
1006-1011.

\bibitem{ejbrody}
{\sc Brody,~E.~J.} (1971) An elementary proof of the Gaussian
dichotomy theorem. {\em Zeitschrift f\"ur Wahrscheinlichkeitstheorie
und Verwandte Gebiete} \textbf{20}, 217-226.

\bibitem{ejbrody2}
{\sc Brody,~E.~J.} (1987) Applications of the Kakutani metric to
real-space renormalization. {\em Physical Review Letters}
\textbf{58}, 179-182.

\bibitem{burbea1}
{\sc Burbea,~J.} (1986) Informative geometry of probability spaces.
{\em Expositiones Mathematicae} \textbf{4}, 347-378.

\bibitem{burbea2}
{\sc Burbea,~J., Oller,~J.~M. \& Reverter,~F.} (2002) Some remarks
on the information geometry of the Gamma distribution. {\em
Communications in Statistics. Theory and Methods} \textbf{31},
1959-1975.

\bibitem{burbea3}
{\sc Burbea,~J. \& Rao,~C.~R.} (1982a) On the convexity of some
divergence measures based on entropy functions. {\em IEEE
Transactions on Information Theory} \textbf{IT-28}, 489-495.

\bibitem{burbea4}
{\sc Burbea,~J. \& Rao,~C.~R.} (1982b) On the convexity of higher
order Jensen differences based on entropy functions. {\em IEEE
Transactions on Information Theory} \textbf{IT-28}, 961-963.

\bibitem{burbea5}
{\sc Burbea,~J. \& Rao,~C.~R.} (1984) Differential metrics in
probability spaces. {\em Probability and Mathematical Statistics}
\textbf{3}, 241-258.

\bibitem{bures}
{\sc Bures,~D.} (1969) An extension of Kakutani's theorem on
infinite product measures to the tensor product of semifinite
$w\sp{\ast}$-algebras. {\em Transactions of the American
Mathematical Society} \textbf{135}, 199-212.

\bibitem{pistone}
{\sc Cena,~A. \& Pistone,~G.} (2007) Exponential statistical
manifold. {\em Annals of the Institute of Statistical Mathematics}
\textbf{59}, 27-56.

\bibitem{chentsov}
{\sc \v{C}encov,~N.~N.} (1982) {\em Statistical Decision Rules and
Optimal Inference}. Translations of Mathematical Monographs
\textbf{53} (Providence: American Mathematical Society). (Originally
published as {\cyr Statisticheskie Rexayuwie Pravila i Optimalbnye
Vyvody}. {\cyr Moskva: Nauka} 1972.)

\bibitem{crooks}
{\sc Crooks,~G.~E.} (2007) Measuring thermodynamic length. {\em
Physical Review Letters} \textbf{99}, 100602.

\bibitem{dawid}
{\sc Dawid,~A.~P.} (1975) Discussion on professor Efron's paper.
{\em The Annals of Statistics} \textbf{3}, 1231-1234.

\bibitem{diosi1}
{\sc Di\'{o}si,~L., Forg\'{a}cs,~G., Luk\'{a}cs,~B. \&
Frisch,~H.~L.} (1984) Metricization of thermodynamic-state space and
the renormalization group. {\em Physical Review} A\textbf{29},
3343-3345.

\bibitem{diosi2}
{\sc Di\'{o}si,~L. \& Luk\'{a}cs,~B.} (1986) Spatial correlations in
diluted gases from the viewpoint of the metric of the thermodynamic
state space. {\em Journal of Chemical Physics} \textbf{84},
5081-5084.

\bibitem{diosi3}
{\sc Di\'{o}si,~L., Luk\'{a}cs,~B \& R\'{a}cz,~A.} (1989) Mapping
the van der Waals state space. {\em Journal of Chemical Physics}
\textbf{91}, 3061-3067.

\bibitem{dolan}
{\sc Dolan,~B.~P.} (1998) Geometry and thermodynamic fluctuations of
the Ising model on a Bethe lattice. {\em Proceedings of the Royal
Society London} A\textbf{454}, 2655-2665.

\bibitem{efron}
{\sc Efron,~B.} (1975) Defining the curvature of a statistical
problem (with applications to second order efficiency). With a
discussion by C.~R.~Rao, Don~A.~Pierce, D.~R.~Cox, D.~V.~Lindley,
Lucien~LeCam, J.~K.~Ghosh, J.~Pfanzagl, Niels~Keiding, A.~P.~Dawid,
Jim~Reeds and with a reply by the author. {\em The Annals of
Statistics} \textbf{3}, 1189-1242.

\bibitem{efron2}
{\sc Efron,~B.} (1978) The geometry of exponential families. {\em
The Annals of Statistics} \textbf{6}, 362-376.

\bibitem{fisher}
{\sc Fisher,~R.~A.} (1925) Theory of statistical estimation. {\em
Proceedings of the Cambridge Philosophical Society} \textbf{122},
700-725.

\bibitem{gibilisco}
{\sc Gibilisco,~P. \& Isola,~T.} (2007) Uncertainty principle and
quantum Fisher information. {\em Annals of the Institute of
Statistical Mathematics} \textbf{59}, 147-159.

\bibitem{gibilisco2}
{\sc Gibilisco,~P., Imparato,~D. \& Isola,~T.} (2007) Uncertainty
principle and quantum Fisher information II. {\em Journal of
Mathematical Physics}. \textbf{48}, 072109.

\bibitem{grasselli}
{\sc Grasselli,~M.~R. \& Streater,~R.~F.} (2001) On the uniqueness
of the Chentsov metric in quantum information geometry. {\em
Infinite Dimensional Analysis, Quantum Probability and Related
Topics} \textbf{4}, 173-182.

\bibitem{havrda}
{\sc Havrda,~J. \& Charv\'at,~F.} (1967) Quantification method of
classification processes. {\em Kybernetika} \textbf{3}, 30-35.

\bibitem{hellinger}
{\sc Hellinger,~E.} (1909) Neue Begr\"undung der Theorie
quadratischer Formen von unendlichvielen Ver\"anderlichen. {\em
Journal f\"ur die reine und angew Mathematik} \textbf{136}, 210-271.

\bibitem{hemmer}
{\sc Hemmer,~P.~C.} (1964) On the van der Waals theory of the
vapour-liquid equilibrium. IV. The pair correlation function and
equation of state for long-range forces. {\em Journal of
Mathematical Physics} \textbf{5}, 75-84.

\bibitem{hemmer1}
{\sc Hemmer,~P.~C., Kac,~M. \& Uhlenbeck,~G.~E.} (1964) On the van
der Waals theory of the vapour-liquid equilibrium. III. Discussion
of the critical region. {\em Journal of Mathematical Physics}
\textbf{5}, 60-74.

\bibitem{ingarden1}
{\sc Ingarden,~R.~S.} (1981), Information geometry in functional
spaces of classical and quantum finite statistical systems. {\em
International Journal of Engineering Science} \textbf{19},
1609-1633.

\bibitem{ingarden2}
{\sc Ingarden,~R.~S. \& Tamassy,~L.} (1993) On parabolic geometry
and irreversible macroscopic time. {\em Reports on Mathematical
Physics} \textbf{32}, 11-33.

\bibitem{Johnston}
{\sc Janke,~W., Johnston,~D.~A. \& Malmini,~R.~P.~K.~C.} (2002)
Information geometry of the Ising model on planar random graphs.
{\em Physical Review} E\textbf{66}, 056119.

\bibitem{janyszek1}
{\sc Janyszek,~H.} (1990) Riemannian geometry and stability of
thermodynamical equilibrium systems. {\em Journal of Physics}
A\textbf{23}, 477-490.

\bibitem{janyszek2}
{\sc Janyszek,~H. \&  Mruga{\l}a,~R.} (1989a) Geometrical structure
of the state space in classical stastical and phenomological
thermodynamics. {\em Reports on Mathematical Physics} \textbf{27},
145-159.

\bibitem{janyszek4}
{\sc Janyszek,~H. \& Mruga{\l}a,~R.} (1989b) Riemannian geometry and
the thermodynamics of model magnetic systems. {\em Physical Review}
A\textbf{39}, 6515-6523.

\bibitem{janyszek3}
{\sc Janyszek,~H. \& Mruga{\l}a,~R.} (1990) Riemannian geometry and
the stability of ideal quantum gases. {\em Journal of Physics}
A\textbf{23}, 467-476.

\bibitem{jeffreys}
{\sc Jeffreys,~H.} (1946) An invariant form for the prior
probability in estimation problems. {\em Proceeding of the Royal
Society London}. A\textbf{186}, 453-461.

\bibitem{jencova}
{\sc Jen\v{c}ov\'{a},~A. \& Petz,~D.} (2006) Sufficiency in quantum
statistical inference: A survey with examples. {\em Infinite
Dimensional Analysis, Quantum Probability and Related Topics}
\textbf{9}, 331-351.

\bibitem{Johnston2}
{\sc Johnston,~D.~A., Janke,~W. \& Kenna,~R.} (2003) Information
geometry, one, two, three (and four). {\em Acta Physica Polonica}
B\textbf{34}, 4923-4937.

\bibitem{kac1}
{\sc Kac,~M.} (1959) On the partition function of a one-dimensional
gas. {\em Physics of Fluids} \textbf{2}, 8-12.

\bibitem{kac2}
{\sc Kac,~M., Uhlenbeck,~G.~E. \& Hemmer,~P.~C.} (1963) On the van
der Waals theory of the vapour-liquid equilibrium. I. Discussion of
a one-dimensional model. {\em Journal of Mathematical Physics}
\textbf{4}, 216-228.

\bibitem{kakutani}
{\sc Kakutani,~S.} (1948) On equivalence of infinite product
measures. {\em Annals of Mathematics} \textbf{49}, 214-224.

\bibitem{kass}
{\sc Kass,~R.} (1989) The geometry of asymptotic inference. With
comments and a rejoinder by the author. {\em Statistical Science}
\textbf{4}, 188-234.

\bibitem{lebowitz}
{\sc Lebowitz,~J.~L. \& Penrose,~O.} (1966) Rigorous treatment of
the van der Waals-Maxwell theory of liquid-vapour transition. {\em
Journal of Mathematical Physics} \textbf{7}, 98-113.

\bibitem{lebowitz2}
{\sc Lebowitz,~J.~L. \& Percus,~J.~K.} (1963) Asymptotic behaviour
of the radial distribution function. {\em Journal of Mathematical
Physics} \textbf{4}, 248-254.

\bibitem{Maha2}
{\sc Mahalanobis,~P.~C.} (1930) On tests and meassures of groups
divergence. {\em Journal of the Asiatic Society of Bengal}
\textbf{26}, 541-588.

\bibitem{Maha3}
{\sc Mahalanobis,~P.~C.} (1936) On the generalised distance in
statistics. {\em Proceedings of the National Institute of Science
India} A\textbf{2}, 49-55.

\bibitem{maybank}
{\sc Maybank,~S.} (2005) The Fisher-Rao metric for projective
transformations of the line. {\em International Journal of Computer
Vision} \textbf{63}, 191-206.

\bibitem{mayer}
{\sc Mayer,~J.~E. \& Mayer,~M.~G.} (1940) {\em Statistical
Mechanics} (New York: John Wiley \& Sons).

\bibitem{mijatovic}
{\sc Mijatovi\'{c},~M., Veselinovi\'{c} \& Trenevski,~K.} (1987)
Differential geometry of equilibrium thermodynamics. {\em Physical
Review} A\textbf{35}, 1863-1867.

\bibitem{mrugala3}
{\sc Mruga{\l}a,~R., Nulton,~J.~D., Sch\"on,~J.~C. \& Salamon,~P.}
(1990) Statistical approach to the geometric structure of
thermodynamics. {\em Physical Review} A\textbf{41}, 3156-3160.

\bibitem{murray93}
{\sc Murray,~M.~K. \& Rice,~J.~W.} (1993) {\it Differential Geometry
and Statistics} (London: Chapman \& Hall).

\bibitem{oconnor}
{\sc O'Connor,~D. \& Stephens,~C.~R.} (1993) Geometry, the
renormalisation group and gravity. In {\em Directions in general
relativity} (ed. B.~L.~Hu, M.~P.~Ryan~Jr \& C.~V.~Vishveshwava).
Proc. 1993 Int. Symp., Maryland, \textbf{1}. (Cambridge: Cambridge
University Press).

\bibitem{peter}
{\sc Peter,~A. \& Rangarajan,~A.} (2006) Shape analysis using the
Fisher-Rao Riemannian metric: unifying shape representation and
deformation. {\em Proceedings of the 3rd IEEE International
Symposium on Biomedical Imaging: Nano to Macro}, 1164-1167.

\bibitem{petz}
{\sc Petz,~D.} (2002) Covariance and Fisher information in quantum
mechanics. {\em Journal of Physics}. A\textbf{35}, 929-939.

\bibitem{petz2}
{\sc Petz,~D. \& Sudar,~C.} (1996) Geometries of quantum states.
{\em Journal of Mathematics Physics}. \textbf{37}, 2662-2673.

\bibitem{rao1}
{\sc Rao,~C.~R.} (1945) Information and the accuracy attainable in
the estimation of statistical parameters. {\em Bulletin of the
Calcutta Mathematical Society} \textbf{37}, 81-91.

\bibitem{rao2}
{\sc Rao,~C.~R.} (1947) The problem of classification and distance
between two populations. {\em Nature} \textbf{159}, 30-31.

\bibitem{rao3}
{\sc Rao,~C.~R.} (1954) On the use and interpretation of distance
functions in statistics. {\em Bulletin de l'Institut International
de Statistique} \textbf{34}, 90-97.

\bibitem{rao4}
{\sc Rao,~C.~R.} (1984) Convexity properties of entropy functions
and analysis of diversity. In {\em Inequalities in statistics and
probability}. Proceeding of the Symposium on Inequalities in
Statistics and Probability. Lincoln, Nebraska 1982. (Ed. Y.~L.~Tong)
Institute of Mathematical Statistics Lecture Notes. Monograph Series
\textbf{5}. (Hayward, CA: Institute of Mathematical Statistics).

\bibitem{ruppeiner0}
{\sc Ruppeiner,~G.} (1979) Thermodynamics: A Riemannian geometric
model. {\em Physical Review} A\textbf{20}, 1608-1613.

\bibitem{ruppeiner1}
{\sc Ruppeiner,~G.} (1990) Thermodynamic curvature of a
one-dimensional fluid. {\em Journal of Chemical Physics}
\textbf{92}, 3700-3709.

\bibitem{ruppeiner2}
{\sc Ruppeiner,~G.} (1991) Riemannian geometric theory of critical
phenomena. {\em Physical Review} A\textbf{44}, 3583-3595.

\bibitem{ruppeiner3}
{\sc Ruppeiner,~G.} (1995) Reimannian geometry in thermodynamic
fluctuation theory. {\em Reviews of Modern Physics} \textbf{67},
605-659.

\bibitem{ruppeiner4}
{\sc Ruppeiner,~G.} (2008) Thermodynamic curvature and phase
transitions in Kerr-Newman black holes. {\em Physical Review}
D\textbf{78}, 024016.

\bibitem{ruppeiner5}
{\sc Ruppeiner,~G. \& Chance,~J.} (1990) Reimannian geometry in
thermodynamic fluctuation theory. {\em Journal of Chemical Physics}
\textbf{92}, 3700-3709.

\bibitem{salamon}
{\sc Salamon,~P. \& Berry,~R.~S.} (1983) Thermodynamic length and
dissipated availability. {\em Physical Review Letters} \textbf{51},
1127-1130.

\bibitem{schlogl}
{\sc Schl\"ogl,~F.} (1985) Thermodynamic metric and stochastic
measures. {\em Zeitschrift f\"ur Physik} B\textbf{59}, 449-454.

\bibitem{streater}
{\sc Streater,~R.~F.} (2004) Duality in quantum information
geometry. {\em Open Systems and Information Dynamics} \textbf{11},
71-77.

\bibitem{tonks}
{\sc Tonks,~L.} (1936) The complete equation of state of one, two,
and three-dimensional gases of hard elastic spheres. {\em Physical
Review} \textbf{50}, 955-963.

\bibitem{tsallis}
{\sc Tsallis,~C.} (1988) Possible generalization of Boltzmann-Gibbs
statistics. {\em Journal of Statistical Physics} \textbf{52},
479-487.

\bibitem{uhlenbeck}
{\sc Uhlenbeck,~G.~E., Hemmer,~P.~C. \& Kac,~M.} (1963) On the van
der Waals theory of the vapour-liquid equilibrium. II. Discussion of
the distribution functions. {\em Journal of Mathematical Physics}
\textbf{4}, 229-247.

\bibitem{uhlmann}
{\sc Uhlmann,~A.} (1996) Spheres and hemispheres as quantum state
space. {\em Journal of Geometry and Physics} \textbf{18}, 76-92.

\bibitem{vanhove}
{\sc Van~Hove,~L.} (1950) Sur l'int\'egral de configuration pour les
syst\`{e}mes de particules \`{a} une dimension. {\em Physica}
\textbf{16}, 137-143.

\bibitem{vankampen}
{\sc Van~Kampen,~N.~G.} (1964) Condensation of a classical gas with
long-range attraction. {\em Physical Review} \textbf{135},
A362-A369.

\bibitem{weinhold}
{\sc Weinhold,~F.} (1975) Metric geometry of equilibrium
thermodynamics. {\em Journal of Chemical Physics} \textbf{63},
2479-2483.

\bibitem{yau}
{\sc Yau,~S.~T.} (1977) Calabi's conjecture and some new results in
algebraic geometry. {\em Proceedings of the National Academy of
Science U.S.A.} \textbf{74}, 1798-1799.

\bibitem{zanardi}
{\sc Zanardi,~P., Giorda,~P. \& Cozzini,~M.} (2007)
Information-theoretic differential geometry of quantum phase
transitions. {\em Physical Review Letters} \textbf{99}, 100603.


\end{thebibliography}
\end{document}